\documentclass[11pt,a4paper]{article}

\def\bu{\delta}

\def\TT{{\cal T}}

\def\WW{{\cal W}}

\def\ZZ{{\cal Z}}

\def\nnu{{\rho}}

\usepackage{color}
\usepackage{jheppub}
\usepackage{amssymb}
\usepackage{amsmath}

\def\z{{\zeta}}

\def\bC{{\bf C}}
\def\bR{{\bf R}}

\def\bZ{{\bf Z}}
\def\Im{\mathop{\rm Im}\nolimits}
\def\Re{\mathop{\rm Re}\nolimits}

\def\th{\mathop{\rm th}\nolimits}
\def\ch{\mathop{\rm ch}\nolimits}
\def\sh{\mathop{\rm sh}\nolimits}

\def\TT{{\cal T}}

\def\WW{{\cal W}}

\def\ZZ{{\cal Z}}
\def\wh{\widehat}
\def\wt{\widetilde}

\def \veps{\varepsilon}

\def\half{{\scriptstyle{1 \over 2}}}
\def\interior#1{\setbox1=\hbox{$#1$}\rlap{$#1$}\kern0.4\wd1\raise1.1\ht1%
\hbox{$\scriptstyle \circ$}}

\def\boxit#1#2{\setbox1=\hbox{\kern#1{#2}\kern#1}%
\dimen1=\ht1 \advance \dimen1 by #1 \dimen2=\dp1 \advance \dimen2 by #1
\setbox1=\hbox{\vrule height\dimen1 depth\dimen2\box1\vrule}%
\setbox1=\vbox{\hrule\box1\hrule}%
\advance \dimen1 by .4pt \ht1=\dimen1 \advance \dimen2 by .4pt \dp1=\dimen2
\box1\relax}
\def\endprf{\raise .5ex\hbox{\boxit{2pt}{\ }}}

\def\wchi{{\wt \chi}}

\def\ifundefined#1{\expandafter\ifx\csname#1\endcsname\relax}

\def\beq{\begin{equation}}
\def\endq{\end{equation}}
\def\beqa{\begin{eqnarray}}
\def\endqa{\end{eqnarray}}

\newcommand\n{\kappa}

\newcommand\Lobad{{\bf H}_{d}}

\newcommand{\sd}{{\bf S}_{d-1}}

\let\UnmodifSec=\section
\renewcommand{\section}{\setcounter{equation}{0}\UnmodifSec}

\newtheorem{theorem}{Theorem}[section]
\newtheorem{remark}{Remark}[section]

\def\z{{\zeta}}
\def\bC{{\bf C}}
\def\bR{{\bf R}}

\def\bZ{{\bf Z}}

\def\Im{\mathop{\rm Im}\nolimits}
\def\Re{\mathop{\rm Re}\nolimits}

\def\th{\mathop{\rm th}\nolimits}
\def\ch{\mathop{\rm ch}\nolimits}
\def\sh{\mathop{\rm sh}\nolimits}

\def\TT{{\cal T}}

\def\WW{{\cal W}}

\def\ZZ{{\cal Z}}
\def\wh{\widehat}
\def\wt{\widetilde}

\def \veps{\varepsilon}

\def\half{{\scriptstyle{1 \over 2}}}
\def\interior#1{\setbox1=\hbox{$#1$}\rlap{$#1$}\kern0.4\wd1\raise1.1\ht1%
\hbox{$\scriptstyle \circ$}}

\def\boxit#1#2{\setbox1=\hbox{\kern#1{#2}\kern#1}%
\dimen1=\ht1 \advance \dimen1 by #1 \dimen2=\dp1 \advance \dimen2 by #1
\setbox1=\hbox{\vrule height\dimen1 depth\dimen2\box1\vrule}%
\setbox1=\vbox{\hrule\box1\hrule}%
\advance \dimen1 by .4pt \ht1=\dimen1 \advance \dimen2 by .4pt \dp1=\dimen2
\box1\relax}
\def\endprf{\raise .5ex\hbox{\boxit{2pt}{\ }}}

\def\wchi{{\wt \chi}}

\def\ifundefined#1{\expandafter\ifx\csname#1\endcsname\relax}

\def\beq{\begin{equation}}
\def\endq{\end{equation}}
\def\beqa{\begin{eqnarray}}
\def\endqa{\end{eqnarray}}

\def\k{\kappa}

\renewcommand{\cosh}{\ch}\renewcommand{\sinh}{\sh}
\renewcommand{\tanh}{\th}

\allowdisplaybreaks[1]

\title{Loops in Anti de Sitter space\footnote{To Jacques Bros and Michel Gaudin. In loving memory.}}

\author[1,2]{Sergio L. Cacciatori}
\author[3]{Henri Epstein}
\author[1,2]{Ugo Moschella}

\affiliation[1]{Universit\`a  dell'Insubria, Dipartimento di Scienza ed Alta Tecnologia,\\
 via Valleggio 11, 22100, Como, Italy}
\affiliation[2]{INFN, via Celoria 16, 20133, Milano, Italy}
\affiliation[3]{IHES, Bures-sur-Yvette, France}

\emailAdd{sergio.cacciatori@uninsubria.it}
\emailAdd{epstein@ihes.fr}
\emailAdd{ugo.moschella@uninsubria.it}

\abstract{We discuss general one and two-loop banana diagrams and one-loop diagrams with external lines with arbitrary masses on the anti de Sitter spacetime by using methods of AdS quantum field theory in the dimensional regularization approach. 
The banana diagrams explicitly computed in this paper are indeed the necessary ingredients for the evaluation of the two-loop effective potential of the Standard Model and can be used to extend the flat space results in 
in presence of a negative cosmological constant. In the one-loop case we also compute the effective potential for an $O(N)$ model in $d=4$ dimension as an explicit function of the cosmological constant $\Lambda$, both exactly and perturbatively up to order $\Lambda$. In the two-loop case we show  the explicit calculation is possible  thanks to a remarkable discrete K\"all\'en-Lehmann formula which we found and proved sometimes ago and whose domain of applicability we extend in the present paper.}

\makeatletter
\gdef\@fpheader{}
\makeatother
\begin{document}

\maketitle

\section{Introduction}
The outcome  of string theory having the largest impact in science is probably  the  AdS/CFT correspondence \cite{maldacena-last,maldacena-first,DeWolfe:2018dkl,witten-adscft}. Though still conjectural, its robustness is evident in several situations, going far beyond the realm of string theory and supersymmetry 
\cite{babington}. Trying to explore and understand the conjecture has generated  an impressive literature dealing with various  geometrical and physical aspects  of (asymptotically) anti-de Sitter (AdS) spacetimes \cite{Aharony}; in particular, AdS  Quantum Field Theory (QFT) has come again to the general attention. 

On the AdS side, the conjecture is often checked only at tree-level: Feynman diagrams  on that manifold are  difficult already at one loop. 
Sometimes,  loop integrals are evaluated through the bootstrap, a strategy  introduced by G. Chew in the seventies of the last century,   abandoned shortly after the successes of QCD and the Standard Model, and revitalized in the last fifteen years \cite{ryz}. 
Bootstrap is essentially based on the  interplay between symmetries, unitarity and causality. However, it is not  completely clear how to implement it outside the context of conformal field theories and difficulties may arise in presence of anomalies or non-invertible symmetries \cite{komatsu}. Combining bootstrap methods in CFT with the AdS/CFT conjecture has proven to be useful to guess the result of loop integrals in the AdS ({\em bulk}) manifold \cite{Pene,Carmi2024,Carmi2021, Carmi2019, drummond, Carmi2018}; there are however other possibilities \cite{sachs,akh}. 

Generally speaking, the absence of a global linear momentum space renders the calculation of Feynman or Schwinger-Keldysh diagrams in  a curved background more difficult than usual;  to avoid that difficulty,  working in position space from scratch is the cleanest and clearest option that  may be tremendously effective also in flat space \cite{Cacciatori2023a}. That is the strategy that we adopt here and in a companion paper \cite{HSUdS}.  By taking  into full account  the analyticity properties of the correlation functions in the complexified universal covering of the real AdS manifold, we show how to compute certain one and two-loop integrals in a simple and direct way, avoiding every detour whatsoever.

Let us focus on one loop integral first: sometimes  things are rendered more difficult than they really are by the habit of writing the two-point functions and the propagators as special instances of the Gaussian hypergeometric function ${}_2F_1$ instead of using the more specific expression in terms Legendre functions  of the second kind ($Q$) for AdS  quantum fields and  Legendre functions of the first kind ($P$) for dS fields.  
This might look just as an irrelevant matter of taste but the contrary happens to be true:  the geometrical peculiarities of either the AdS or the dS manifolds that are well captured by Legendre functions  remain hidden in the generic hypergeometric expression.  The superiority of using $Q$ instead of ${}_2F_1$ is  evident in our first computation of  the 1-loop diagram  resulting in Eq. (\ref{bubads}). This result in turn opens the way to the calculation of two-loop banana integrals for three independent AdS scalar fields having different values of their respective mass parameters. We present also a second derivation of Eq. (\ref{bubads}) as a special instance of the new formula (\ref{formu3}) which is generally valid in any curved Euclidean background.

The literature on AdS {\em two-loop} integrals is scarce, see e.g. \cite{drummond,Roiban,sachs}. In general the relevant  articles do not contain explicit formulas: some of them calculate corrections to the dispersion laws but do so determining only the divergent part \cite{Sundin}. 

The explicit calculation in the present paper is made possible by the AdS K\"all\'en-Lehmann expansion formula which two of us and collaborators proved sometimes ago \cite[Eqs 7.8 and 7.9]{begmp2} (Eqs. (\ref{x.15})  and (\ref{kl}) in this paper). A  formula identical to our own appeared without an explicit proof in a later paper the same year \cite[Eq. 53]{fitzpatrick}\footnote{Numbered as Eq. 52 in the first version arXiv:1111.6972v1 of the preprint.}.

Remarkably, this beautifully simple formula is not an integral but a discrete sum.  One could have expected this for the uncovered AdS manifold, due to its time periodicity; however, the discreteness of the formula is not affected by moving to the universal covering. 
Here we extend the validity of  
K\"all\'en-Lehmann expansion to more general conditions in Theorem \ref{k-l4} : the  formula remains of course the same as in \cite{begmp2}, only the domain of applicability becomes wider w.r.t. the original proof given in \cite{begmp2}. Then we use it together with our 1-loop calculation to write explicit formulae for the two-loop banana integral with three independent masses  in arbitrary (complex) dimension $d$  (Eqs. (\ref{bub3bis}), (\ref{AA1}) and (\ref{AA2})). Obtaining these formulae  would be quite challenging using other methods without further assumptions (for example, without assuming the validity of the AdS/CFT conjecture and/or doing suitable guesses in bootstrap methods).  The K\"all\'en-Lehmann expansion renders the  calculation is next to elementary by linearization. 

We would like to stress also that computing these integrals goes well beyond  the aim of testing  the power of the methods used in the present paper: these are exactly the diagrams necessary to compute the two-loop effective potential for the Standard Model and can be used to generalize the results in 
\cite{Ford:1992pn} in presence of a negative cosmological constant. The simple explicit formulas we have found are new and we think that they can lead to some simplifications also in the flat limit.

The plan of the paper is as follows:  in Sect. \ref{sec3} we recall a few preliminary notions about the geometry of AdS and about AdS QFT.  
In Sect. \ref{anal} we discuss in more detail the analyticity properties of the scalar two-point functions and recall how to write the propagators in terms of Legendre functions of the second kind. A simple application of this representation  is presented in Sect. \ref{sec4} where we compute the 1-loop banana integral with two arbitrary masses in any complex dimensions by using the Wronskian associated to the Legendre differential equation.  

The bubble is reconsidered again in Sect. \ref{sec5a} where we prove the general formula (\ref{formu3}) valid  in any (curved) Euclidean background. In the same vein we  determine also a formula the triangular loop that may be easily generalized to $N$-line case.

Sect. \ref{6} contains a curious fact we found  on the relation betweeen bulk diagram vs Witten diagrams which is worth of further investigation. We show that in at least one possibly pathological case, Witten diagrams are not the boundary limit of the correspding bulk diagram.

In Sect. \ref{kal} we discuss the K\"all\'en-Lehmann formula for AdS scalar quantum fields with two distinct masses and we extend the original proof given in \cite{begmp2} to more general conditions. We also discuss instructive  elementary deductions of the same  formula in odd dimensions $d=-1,\ d=1,\ d=3,\ d=5$. 
In Sects. \ref{8} and \ref{wmelon} we use the KL expansion formula to compute  explicit  expressions for the 1-loop tadpole and the one loop sunset and  for the 2-loop banana integral with arbitrary masses and dimensions.

In the final Sect. \ref{sec5} we apply the   results of Sect. \ref{sec4} to compute the 1-loop effective potential for an  AdS $O(N)$ scalar field model in a four dimensions.  We provide an exact expression (at 1-loop) of the potential  as a function of the cosmological constant $\Lambda$ and  also an expansion in $\sqrt{-\Lambda}$  to  order $\Lambda$; in the flat limit $\Lambda\to 0$  the standard flat potential is correctly reproduced. The latter fact is indeed nontrivial.
The text is followed by two Appendices. In the first  we provide some further details about the usage of Legendre functions of first and second kind in dS and AdS QFT's and  also clarify some points that have been discussed in the recent literature. In the second we give the details about the proof of the expansion theorem of \cite{begmp2} in general conditions.


\section{Preliminaries}\label{sec3}

\subsection{Geometry}
The $d$-dimensional real AdS spacetime
with radius $R>0$ may be visualized as the manifold
\begin{align}
    AdS_d = \{x \in \bR^{d+1}\ :\ x^2= x\cdot x = R^2\}
\label{s.1}
\end{align}
where the scalar product $x\cdot x$ is intended in the sense of the ambient space $ \bR^{d+1}$  with two timelike directions and metric mostly minus as follows:
\begin{align}
   x\cdot y  = x^0 y^0-x^1y^1-\ldots-x^{d-1}y^{d-1} +x^dy^d. 
\label{metricads}\end{align}
The complexification of the AdS manifold is defined analogously 
\beq
AdS_d^{(c)} = \{z =x+iy \in \bC^{d+1}\ :\ z^2 = R^2\};
\label{complexmanifold}\endq
 $z\in AdS^{(c)}$ if and only if  $x^2 -y^2=R^2$ and $x\cdot y = 0$, i.e. the real and imaginary parts of $z$ are orthogonal w.r.t. the scalar product (\ref{metricads}). 
The symmetry group of the anti de Sitter spacetime is is the pseudo-orthogonal group of the ambient space $SO(2,d-1)$. This group may also be regarded as the the conformal group of transformations of the boundary, represented as the null cone of the ambient space
\begin{align}
 C_d = \{\xi \in \bR^{d+1}\ :\ \xi^2= \xi\cdot \xi = 0\}.
\label{ncone}
\end{align}
This simple geometrical fact lies at basis of the AdS/CFT correspondence. The null cone of the ambient space plays also the role of 
giving a causal order to the AdS spacetime  which is however only {\em local}, due to the existence of closed timelike curves;
two events are spacelike separated if
\begin{equation}
(x_1-x_2)^2= 2 -2x_1\cdot x_2 <0. \label{causalads}
\end{equation}

While  $AdS_d^{(c)}$ is simply connected\footnote{This may be seen by changing $z^\mu$
to $iz^\mu$ for $ \mu= 1,\ldots,d-1,$
the complex AdS manifold becomes 
the complex unit sphere in $\bC^{d+1}$, which has the
same homotopy type as the real unit sphere $S_d$.
It follows that for $d \ge2$ the {covering space} of $AdS_d^{(c)}$ 
is $ AdS_d^{(c)}$ itself.},
the real manifold  $AdS_d$ is not and  admits a nontrivial 
universal {covering space}  
$\wt {AdS}_d$. The covering manifold is globally causal but remains non-globally hyperbolic, because of the boundary at spacelike infinity.
However, even though  time-loops
are not present in the covering space,   there remains an unavoidable trace of the periodicity of the uncovered (otherwise called pure or true) AdS manifold:  timelike geodesics issued at any point of $\wt {AdS}_d$ focus infinitely many times every half period, exactly as it happen  on the true AdS manifold. 
This fact will be conspicuous in Sect. \ref{kal} where we discuss the K\"all\'en-Lehmann representation of the product of two scalar two-point functions with mass parameters  $\lambda$ and $\nu$, in position space with fixed end-points (sometimes called the bubble diagram).

\subsection*{Coordinates}

Except for the considerations regarding the flat limit  we shall always take  $R=1$ throughout
this paper.
A concrete way of representing 
both  ${AdS}_d$  
{and its covering} $\wt {AdS}_d$ by  global coordinates is to introduce   
the diffeomorphism $\chi$ of ${ S}_1\times \bR^{d-1}$ onto ${AdS}_d$ given by
\begin{equation}
x(t, \vec{x}) =  
(\sqrt{1+\vec{x}^2}\,\cos t,\ \vec{x},\ \sqrt{1+\vec{x}^2}\,\sin t)
\label{2.11}\end{equation}
where ${S}_1$ is identified to $\bR/2\pi \bZ$. The diffeomorphism $\wchi$, defined by lifting   
$\chi$ on the covering $\bR^{d}$ of  
${ S}_1\times \bR^{d-1}$
provides a global coordinate system on $\wt {AdS}_d$.

It is possible to identify in the complex manifold ${AdS}_d^{(c)}$ an analog of 
the Euclidean subspace of the complex Minkowski
spacetime:
here we choose the connected real submanifold 
$\Lobad$
of ${AdS}_d^{(c)}$  -- that we call the {Euclidean anti de Sitter spacetime} --
defined by 
\beq
\Lobad = \{z \in {AdS}_d^{(c)} \ \  z(y) =  (y^0,\ldots, y^{d-1},i y^d ),\ y^\mu\in {\bf R }, \ \ y^0 >0\}.
\label{s.1}\endq
This is indeed a  Lobachevsky
space modeled as the upper sheet of a two-sheeted hyperboloid embedded in a $(d+1)$-dimensional Minkowski spacetime:
\begin{equation}
H_d  = \{y\in{\bf R}^{d+1}: \ \  {y^0}^2-{y^1}^2- \dots- {y^d}^2=R^2,\ \  y^{0}> \label{lobad}
0\}.
\end{equation}
Coordinates for $\Lobad$ are  obtained by changing $t$ into $ is$ in Eq. (\ref{2.11}); this yields the following
parametrization: 
\begin{equation}
z(i s, \vec{x}) =  
(\sqrt{1+\vec{x}^2}\,{\cosh}\  s,\ \vec{x},\ i \sqrt{1+\vec{x}^2}\, \sinh s)
\label{parameucl}\end{equation}
There also exists an
extension  of $\wchi$ 
 to ${\bf C}\times \bR^{d-1}$, whose image is   
a partial complexification of the real covering  manifold 
$\wt {AdS}_d$; the so-obtained complexified covering contains
the same Euclidean spacetime      $\Lobad$. 

We will make use of  the following  spherical parametrization of
$H_d$:
\begin{equation}
y(u,\omega)=  (u, \omega^1 \sqrt{u^2 - 1},\ldots, \omega^d\sqrt{u^2 - 1}) 
\end{equation}
where $u\geq 1$ and ${ \omega}$ are coordinates on the sphere $\sd$; in these coordinates the Lorentz-invariant
measure $ dy$ is written
\begin{equation}
\sqrt g \, dy = (u^2-1)^{\frac{d-2}2}du d\omega \label{measureinv}
\end{equation}
where $d\omega$ denotes the rotation-invariant measure on the
sphere ${{S}}_{d-1}$ normalized as follows
\begin{equation}
\omega_{d}=\int_{{{S}}_{d-1}}d\omega  = \frac{2\pi^\frac d2}{\Gamma\left(\frac d2\right)}.
\label{norms}
\end{equation}

\subsection*{Quantum fields and the energy spectrum  condition}
A general approach to AdS QFT  has been discussed in \cite{bem,bemb}. Here we summarize only the main consequences of the spectral condition; we refer the reader to \cite{bem,bemb} for  more details and other  results.  

We focus on a scalar quantum field on $\wt{AdS}_{d}$ which is fully determined by the set of its 
$n$-point vacuum expectation values:
\begin{equation}
{\cal W}_{n}(x_1, \ldots x_n)
=\langle \Omega, \phi(x_1) \ldots \phi(x_n) \Omega \rangle.
\label{npoint}
\end{equation}
The field 
can be restricted to the uncovered spacetime 
$AdS_{d}$  if and only if it  $2 \pi-$periodic in the time-parameter $t$.
The properties of AdS-invariance, positive-definiteness 
and hermiticity are
formulated exactly as for Minkowskian scalar QFT \cite{bem,Streater} 
and we do not spell them out.

Even though the AdS manifold is not globally hyperbolic,    it is however natural to ask for {\em local commutativity}  by requiring that the fields commute at spacelike separation:  $$
{\cal W}(x_1,\ldots,x_i,x_{i+1},\ldots,x_n) =  {\cal W}(x_1,\ldots,x_{i+1},x_i,\ldots,x_n)$$ 
\begin{equation}
\makebox { \em for all  \ } x_i,x_{i+1}\makebox { \em space-like separated in the sense of\ }\wt {AdS}_{d}.
\end{equation}
For theories well-defined in the uncovered 
$AdS_{d}$ itself,  this amounts at least to  commutativity under the  condition
$(x_i-x_{i+1})^2 <0$.

\vskip 10 pt

The above properties are more or less obvious but do not select a unique solution for a given model and some extra criterion is needed.  
Since the very beginning  of AdS quantum field theory  the  correlation functions and the propagators have been constructed and selected by studying the  behaviour of the modes at the boundary located at spacelike infinity \cite{avis}. This is very natural 
from the viewpoint of the Cauchy problem  because of  the lack of global hyperbolicity.

On the other hand, the fact that the field equations do not uniquely select the propagators is  not a peculiarity of AdS quantum fields theory. 
For infinite quantum systems there are always infinitely many inequivalent representations of  the commutation relations and to select a physically meaningful quantization one always needs to impose some extra physical requirement. This is normally done in terms of a {\em spectral condition},  i.e. a  condition on the energy spectrum of the states of the theory, and we act in this way also in the AdS case. Whatever is the  choice it will  also silently  select the behaviour at spacelike infinity of the modes and the correlators.

Since the parameter of the (covering) group of  rotations in the $(0,d)-$plane
is interpreted as a time variable, we require  that the corresponding generator $M_{0d}$
be represented in the Hilbert space of the theory  by a self-adjoint  operator whose spectrum is
bounded from below. 
By a standard Laplace transform argument \cite{Streater}
in the
corresponding time-variables, this requirement is equivalent to the following analyticity property
of the $n$-point functions:
\\[10pt]
{\bf Spectral condition:}
Each tempered distribution 
\begin{equation}
    {\cal W}_n(x_1(t_1,\vec x_1),...,x_n(t_n,\vec x_n))
\end{equation}
is the boundary value of a
holomorphic function ${W}_n (z_1,...,z_n)$ which is defined in a complex neighborhood
of the set 
\begin{equation}
\left\{z=(z_1,...,z_n); z_j  \in {\wt {AdS}_{d}^{(c)}};
 z_j = z_j(\tau_j, \vec x_j);
\  \Im {\tau_1} <\Im{\tau_2} <\cdots <\Im{\tau_n}\right\}.
\end{equation}
As a by-product, the {\em Schwinger functions}  \begin{equation}
    {S}_n(x_1(is_1,\vec x_1),...,x_n(is_n,\vec x_n))
\end{equation}
are well-defined and are obtained as  the restrictions of the holomorphic functions ${W}_n$'s to the Euclidean submanifold
\begin{equation}
 \left\{z=(z_1,...,z_n), z_j\in \Lobad, \  s_1 <s_2 <\cdots
<s_n\right\}
\end{equation}
{at non coinciding points. Sometimes natural distributional
extensions exist: this is the case if the singularities of the
${\cal W}_n$ are weak, for example in lower dimensions or in the presence
of regularization such as dimensional regularization.}

\section{The analytic structure of two-point functions}

\label{anal}
From now on we consider only fields 
completely determined by their two-point
function ${\cal W}(x_1, x_2)$ as we do in perturbation theory i.e. we assume that the truncated $n$-point functions vanish. 

There are two distinguished complex domains \cite{bem, bemb} of
$AdS^{(c)}_{d}$, invariant
 under real AdS transformations, which are of crucial importance for a full
understanding of the structures associated with two-point functions:
$$ {\cal Z}_{1+}  =\{ z  = x+iy  \in AdS^{(c)}_{d}; \, y^2>0,\, \epsilon(z) = +1\},$$
\begin{equation}
{\cal Z}_{1-}  =\{ z = x+iy  \in AdS^{(c)}_{d}; \, y^2>0,\, \epsilon(z) = -1\},
\label{Tubes}
\end{equation}
where
\begin{equation}
\epsilon(z) =  \mbox{sign}  (y^0 x^{d}- x^0 y^{d}).
\end{equation}
${\cal Z}_{1+}$ and ${\cal Z}_{1-}$  are the AdS analogues of
the usual forward and backward tubes
of complex Minkowski spacetime, obtained in correspondence with
the energy-momentum spectrum condition \cite{Streater}.
{ The spaces ${\cal Z}_{1\pm}$ and $AdS_d$ have the
same homotopy type.}
Their universal coverings are denoted $\wt{\cal Z}_{1\pm}$.
The AdS spectral condition implies that   a general
two-point function  satisfies the following \cite{bem}
\\ [5pt]
\noindent 
{ \bf Normal analyticity condition for two-point functions:}
{\em $\mathcal W\left( x_1,x_2\right)$
is the boundary value of a  function $ W\left(
z_1, z_2\right) $ holomorphic in the  domain
$\wt{\cal Z}_{1-}\times \wt{\cal Z}_{1+}$} 
\begin{align}
&\WW(x_1, x_2) = (\Omega,\ \phi(x_1)\phi(x_2)\,\Omega) =
\lim_{\begin{array}{c}
\scriptstyle z_1 \in \wt{\cal Z}_{1-},\ \ z_2 \in \wt{\cal Z}_{1+}\\
\scriptstyle z_1\rightarrow x_1,\ \ z_2\rightarrow x_2 \end{array}}
W(z_1,z_2)\ , \label{spp}
\end{align}
AdS invariance then implies that to 
$W(z_1,z_2)$ there corresponds a function $w(\zeta)$
of a single complex variable $\zeta$ that can be identified with
$z_1\cdot z_2$ when  $z_1$ and $z_2$ are both in  ${AdS}^{(c)}_{d+1}$;  the function $  w(\zeta) $ is called {\em the reduced two-point function}.
Complex AdS invariance and normal analyticity   imply the following \\[10pt]
{\bf  Maximal analyticity property}:
the reduced two-point function {\em $w(\zeta)$ extends to a function
analytic in the  covering $\wt\Delta_1$ of the
cut-plane 
\begin{equation}\Delta_1 = \{{\bf C} \setminus[-1,1]\}.
\end{equation}
For theories periodic in the time coordinate
 $w(\zeta)$ is  analytic in $\Delta_1$.} For later use we introduce also 
 \beq
 \wh \Delta_1 = \Delta_1 \cup \{\infty\}\ ,
 \endq
 Therefore, the two-point function of  any field
satisfying locality AdS invariance and the spectral condition 
enjoys  {\em maximal analyticity}, 
as it happens in the Minkowski \cite{Streater}
and de Sitter  cases \cite{bm}.
Maximal analyticity, in turn,  
completely  determines the  two-point functions for Klein-Gordon fields and, as a consequence, also selects the boundary behaviour of the modes.

\subsection{Klein-Gordon fields}
Klein-Gordon fields display
the simplest example of the previous analytic structure. For a given mass $m$ the two-point function  ${\cal W}(x_1,x_2)$ must satisfy the equation 
\begin{equation}
(\square_{x_i} + m^2) {\cal W}(x_1,x_2) = 0, \ \ i=1,2,\label{kg}
\end{equation}
w.r.t. both variables, 
where $\square_{x_i}$ is the Laplace-Beltrami operator relative to the AdS metric. 
Such two-point functions are labelled by the (complex) dimension $d$ and a (complex) parameter $\nu$
as follows
\begin{align}
&W^{(AdS)}_\nu(z_1,z_2)=w^d_{\nu}(\zeta) =
{1 \over (2\pi)^{d\over 2}} (\zeta^2-1)^{-\frac{d-2}4}
e^{-i\pi{d-2\over 2}} Q_{-{1\over 2}+\nu}^{d-2\over 2}(\zeta)=
\label{s.9}\\
&=
{\Gamma \left ({d-1\over 2} \right )\over
2\pi^{d+1\over 2}} D_{\nu-\frac{d-1}{2}}^{d-1\over 2}(\zeta)\label{tpads} = \\ & =\frac{ \Gamma
   \left(\frac{d-1}{2}+\nu \right)  \,
  }{ 2\pi ^{\frac{d-1}{2}}  (2\zeta)^{ \frac{d-1}{2}+\nu }\Gamma (\nu +1)}  \ {}_2F_1\left(\frac{d-1}{4}+\frac{\nu }{2},\frac{d+1}{4}+\frac{\nu }{2};\nu
   +1;\frac{1}{\zeta^2}\right)
\label{kgtp}\end{align}
where the various parameters are related as follows:
\begin{equation}
     m^2 = \nu^2 -\frac{(d-1)^2}4.
\end{equation}
At the r.h.s. of  (\ref{s.9})    $Q_{\beta}^{\alpha}(\zeta)$ denotes an associated Legendre function of the second kind \cite{bateman}; at the r.h.s. of Eq. (\ref{tpads})  
$ D_{\sigma}^{\delta}(\zeta)$ is  an ultraspherical (Gegenbauer)  
function of the second kind \cite{szego}; they are  meromorphic functions of $\alpha$ and  $\beta$ and, respectively,  $\delta$ and $\sigma$ \cite{bateman,szego}. As functions of the complex variable $\zeta$, they are analytic in the cut-plane
\begin{equation}\Delta_2 = \{{\bf C} \setminus[-\infty,1]\}.
\end{equation}
Their definitions  in terms of the Gauss hypergeometric function is provided by  the last equality (\ref{kgtp}). { Note in particular that 
$(\z)^{\sigma+2\delta}D_{\sigma}^{\delta}(\zeta)$ 
and $(\z)^{\n+{d-1\over 2}}w^d_{\n}(\zeta)$ are even and analytic in
$\wh \Delta_1$}.

Since for each value of the mass squared there are two possible two-point functions, the question arises whether they are both acceptable or not.
The spectral condition gives the answer (Breitenlohner and Freedmann phenomenon) \cite{bf}: the  two-point functions give rise to a representation of of $M_{0d}$ having  positive spectrum only when $\nu$ is real and bigger than  $-1$. There  are two possible cases  
\vskip 10pt

1. for $\nu > 1$ the spectrum condition  (\ref{spp}) uniquely select one field theory for each given value of mass parameter $\nu$;
\vskip 10pt

2. for $|\nu| < 1$ there are two acceptable theories for each given
mass. The difference
between the two theories is in their
large distance behavior; more precisely, in view of  \cite[Eq. (3.3.1.4)]{bateman} one has that 
\begin{equation}
w^d_{-\nu} (\zeta)  = w^d_{\nu} (\zeta)  + \frac{\sin
\pi\nu \, \Gamma\left(\frac {d-1} 2-\nu\right) \Gamma
\left(\frac {d-1} 2 +\nu\right) }{2(2\pi)^{\frac{d}{2}}} \, (\zeta^2-1)^{-\frac {d-2}4}P^{-\frac
{d-2}2}_{-\frac 1 2 -\nu}(\zeta) . \label{pop}
\end{equation}
The last term in this relation is {\em regular on the cut}
$\zeta \in [-1,1]$ and therefore does not contribute to the
commutator. By consequence the two theories represent the same
algebra of local observables at short distances. But since the
second term at the r.h.s. grows the faster the larger is
$|\nu|$ (see \cite[Eqs. (3.9.2))]{bateman} the two theories have 
drastically different  long range behaviors.
\vskip 10 pt

The Schwinger function (otherwise called the Euclidean  propagator) is the restriction of the maximally analytic  two-point function  to the Euclidean Lobachevsky manifold. Choosing the points in Eq. (\ref{s.9}) as follows 
\begin{align}
& z_0= \left(
1 ,
0 ,
\ldots
0,
0 
\right),
 \ \ \ \ \ z(u,\omega)= \left(
u,
\omega^1 \sqrt{u^2-1},
\ldots,
\omega^{d-1} \sqrt{u^2-1},
i \omega^{d} \sqrt{u^2-1}
\right), \ \ \  u>1 \label{uuu}
\end{align}
so that $ \zeta=  z_0\cdot 
z (u,\omega) = u>1$, we write  the propagator as 
\begin{equation}
 G^{(AdS)d}_{\nu}(z_0\cdot 
z (u,\omega))= G^{(AdS)d}_{\nu}(u) = G^d_{\nu}(u)=w_\nu(u) = \frac
{e^{-i\pi\frac {d-2}2}}{(2\pi)^{\frac{d}2}} (u^2-1)^{-\frac
{d-2}4} Q^{\frac {d-2}2}_{-\frac 1 2+\nu}(u). \label{kgtps}
\end{equation}

\begin{remark}
    {\em  We stress  again that  forgetting, as is  frequently done in the literature, that the two-point function is a Legendre function  and not just a much less specific (although equivalent) hypergeometric   function, implicates a significant loss of information and a consequent increase in the  difficulties in  solving even simple problems.

This oblivion is even more curious when one thinks that in one of the first (if not the very first) papers dealing with AdS QFT  \cite{fronsdal} the correlation functions are written exclusively in terms of  Legendre functions $Q$.}
\end{remark}

%

\section{Banana integrals: 1-loop}\label{sec4}
\label{1loop}
After this preparation we can now write the  $n$-loop banana integral on the Lobachevsky Euclidean manifold $\Lobad$ with $n+1$ lines:
\begin{eqnarray}
I_{n+1}(\nu_1,\ldots,\nu_{n+1},d)= \int_{\Lobad} G^{d}_{\nu_1}(x\cdot z)  G^{d}_{\nu_2}(x\cdot z) \ldots  G^{d}_{\nu_{n+1}}(x\cdot z )\sqrt{g(z)}\, dz ,  \label{diagram}\end{eqnarray}
where $y$ varies on  $H_d$ and $x$ is a fixed reference point. The above definition has to be intended as a dimensional regularization of an expression that in general is divergent. 

Using the coordinates (\ref{uuu}) and integrating over the angles  (\ref{diagram}) reduces to 
\begin{eqnarray}
I_{n+1}(\nu_1,\ldots,\nu_{n+1},d)=
  \frac {2\pi ^{\frac {d}{2}}}{\Gamma \left({\frac {d}{2}}\right)} \int _1^\infty G^{d}_{\nu_1}(u )  G^{d}_{\nu_2}(u )  \ldots G^{d}_{\nu_{n+1}}(u )(u^2-1)^{\frac{d-2}2}du.\label{bub}
\end{eqnarray}
The zero-loop case gives 
\begin{eqnarray}
I_{1}(\nu,d)=
  \frac {2 ^{\frac {2 -d}{2}} }{\Gamma \left({\frac {d}{2}}\right)} \int _1^\infty 
{ }{e^{-i\pi\frac {d-2}2}} Q^{\frac {d-2}2}_{-\frac 1 2+\nu}(u) (u^2-1)^{\frac{d-2}4}du. \label{0loop}) 
\end{eqnarray}
The value of this integral might be deduced from \cite[3.12 (29), p. 172] {bateman}  also reproduced in \cite[ET\ II324 (23) p. 809]{grad}, but,  unfortunately, those formulae are flawed by the same typo; we take this opportunity to write here the correct general formula:
\begin{align}
    \int_1^\infty Q^\mu_\nu(u) (u^2-1)^{\lambda-1}du=\frac {\Gamma\Big(\frac 12+\frac \nu2+\frac \mu2 \Big) \Gamma\Big(1-\lambda+\frac \nu2\Big) \Gamma\Big(\lambda+\frac \mu2 \Big) \Gamma\Big(\lambda-\frac \mu2 \Big)}{2^{2-\mu}\Gamma\Big(1+\frac \nu2-\frac \mu2 \Big)\Gamma\Big(\frac 12+\lambda+\frac \nu2 ,\Big)},
\end{align}
valid for $|\mu|<2\Re{\lambda}<\Re{\nu}+2$; this gives the same result as in flat space:
\begin{equation}
I_1(\nu,d)   = \frac 1{\nu^2-\frac {(d-1)^2}4}= \frac 1{m^2}. \label{I1}
\end{equation}

Let us  now  focus on the the already nontrivial two-line case. The above expression looks particularly clean when expressed in terms of the Legendre functions $Q$:
\begin{equation}
I_2(\lambda,\nu,d)   = \int_{H_d} G^{d}_{\lambda}(x\cdot z)  G^{d}_{\nu}(x\cdot z) \sqrt {g(z)}\, dz  
  = \frac { e^{-i\pi( {d-2})}}{ 2^{d-1}\pi ^{\frac {d}{2}}\Gamma \left({\frac {d}{2}}\right)} \int _1^\infty  Q^{\frac {d-2}2}_{-\frac 1 2+\lambda}(u) Q^{\frac {d-2}2}_{-\frac 1 2+\nu}(u)du.\label{bub}
\end{equation}
As simple as it it may look, the integral at the r.h.s. of Eq. (\ref{bub}) is not available anywhere in the literature accessible to us.

Here  the advantage of expressing the propagator in terms of Legendre functions is manifest, as the above integral can be evaluated by using the Wronskian relations associated with  Legendre's differential equation.

Recall that the function $Q$ is holomorphic in the domain
$\Delta_2$; the function  $z\mapsto (z^2-1)^\half$ is also understood as  holomorphic in $\Delta_2$ and 
equal to $|u^2-1|^\half$ when $z=u>1$.
The known  recursion relations 
\begin{align}
& (1-z^2){dQ_\nnu^\mu(z)\over dz} =
-\nnu zQ_\nnu^\mu(z) + (\nnu+\mu)Q_{\nnu-1}^\mu(z)\ ,
\label{c.30}\\
& Q_{\nnu-1}^\mu(z) =
zQ_\nnu^\mu(z) -(\nnu-\mu+1)(z^2-1)^\half Q_\nnu^{\mu-1}(z)\ ,
\label{c.31}\end{align}
(see \cite[3.8 (10), (5) p. 161]{bateman}) give
\beq
(1-z^2){dQ_\nnu^\mu(z)\over dz} =
\mu zQ_\nnu^\mu(z) -(\nnu+\mu)(\nnu-\mu+1)(z^2-1)^\half Q_\nnu^{\mu-1}(z)]\ .
\label{c.32}\endq
These identities allow us to evaluate the relevant integral over  any arc,  with extremities $a$ and $b$, entirely contained in $\Delta_2$: 
\begin{align}
&\int_a^b Q_\nnu^\mu(z)Q_\sigma^\mu(z)(\nnu-\sigma)(\sigma+\nnu+1)dz =\Big [ -(z^2-1)^\half(\sigma+\mu)(\sigma-\mu+1)Q_\nnu^\mu(z)Q_\sigma^{\mu-1}(z)\cr
&+(z^2-1)^\half(\nnu+\mu)(\nnu-\mu+1)Q_\nnu^{\mu-1}(z)Q_\sigma^\mu(z) \Big ]_a^b\ .
\label{c.33}\end{align}
We need also the asymptotic behaviour of the Legendre function of the second kind at $z \sim 1$  and $z \rightarrow \infty$ \cite[3.9.2 (21), (5), (6) pp 163-164]{bateman}:
\begin{align}
& {\rm as}\ \ z \rightarrow +\infty,\ \ Q_\nnu^\mu(z) \sim {\rm const.\ }
z^{-\nnu-1}\ ;   
\label{wq.20}\\
&   {\rm as}\ \ z \rightarrow 1,\ \  {\rm if\ }\Re \mu >0,\ 
 \ Q_\nnu^\mu(z) \sim 
e^{i\pi\mu} 2^{{\mu\over 2}-1}\Gamma(\mu)(z-1)^{-{\mu\over 2}}\ ;   
\label{wq.25}\\
&  {\rm as}\ \ z \rightarrow 1,\ \  {\rm if\ }\Re \mu < 0,\ \ Q_\nnu^\mu(z) \sim
{e^{i\pi\mu} 2^{-{\mu\over 2}-1}\Gamma(-\mu)\Gamma(\nnu+\mu+1)(z-1)^{{\mu\over 2}}
\over \Gamma(\nnu-\mu+1)}\ .   
\label{wq.30}\end{align}
We deduce that the integral
\beq
\int_1^\infty Q_\nnu^\mu(z) Q_\sigma^\mu(z) dz\ 
\label{wq.10}
\endq
converges if
$\Re(\nnu+\sigma) > -1$ and $|\Re \mu| < 1$. 
Let us suppose  that 
$0< \Re \mu <1$. 
It follows that 
\begin{align}
\int_1^\infty Q_\nnu^\mu(z) Q_\sigma^\mu(z)\, dz = 
{e^{2i\pi\mu}\Gamma(\mu)\Gamma(1-\mu)\over
2(\nnu-\sigma)(\sigma+\nnu+1)}
\Bigg [ {\Gamma(\nnu+\mu+1)\over  \Gamma(\nnu-\mu+1)}
-{\Gamma(\sigma+\mu+1)\over  \Gamma(\sigma-\mu+1)} \Bigg ]\ .
\label{wq.50}\end{align}
This equation remains valid, by analytic continuation, when $|\Re \mu| <1$
and $\Re(\nnu+\sigma)  > -1$ (note that the term in brackets vanishes for $\mu=0$).
Letting $\nnu$ tend to
$\sigma$ (\ref{wq.50})  tends to
\begin{align}
\int_1^\infty Q_\sigma^\mu(z) Q_\sigma^\mu(z)\, dz = 
{e^{2i\pi\mu}\Gamma(\mu)\Gamma(1-\mu)\Gamma(\sigma+\mu+1)\over
2(2\sigma+1)\Gamma(\sigma-\mu+1)}\Big [\psi(\sigma+\mu+1)-\psi(\sigma-\mu+1)
\Big ]\ .
\label{wq.60}\end{align}
Again this equation is valid when $|\Re \mu| <1$ and $\Re 2\sigma > -1$ but the r.h.s. can be continued outside of this region.

\vskip 10pt

In the end, the above argument  provides 
a nice formula for the 1-loop banana integral with two independent masses; here it is:
\begin{equation}
I_2(\lambda,\nu,d)  
=\frac{ \Gamma \left(1-\frac{d}{2}\right) }{2^{d} \pi ^{\frac{d}{2}}(\lambda^2 -\nu^2 ) }\
\left(\frac{\Gamma
   \left(\frac{d-1}{2}+\nu \right)}{\Gamma
   \left(\frac{3-d}{2}+\nu
   \right)}-\frac{\Gamma
   \left(\frac{d-1}{2}+\lambda
   \right)}{\Gamma
   \left(\frac{3-d}{2}+\lambda
   \right)}\right). \label{bubads}
   \end{equation}

\vskip 10 pt 
\noindent A few remarks are in order: 
\begin{enumerate}
\item At $d=2$ we may extract  from the general formula the following  finite result:
\begin{equation}
I_2(\lambda,\nu,2) =
\frac{\psi \left(\lambda +\frac{1}{2}\right)-\psi
   \left(\nu +\frac{1}{2}\right)}{2 \pi  \lambda
   ^2-2 \pi  \nu ^2}.
\end{equation}
\item  In odd spacetime dimension the formula becomes
very simple; for instance at $d=1,3$
\begin{equation}
I_2(\lambda,\nu,1) = \frac{1}{2 \lambda ^2 \nu +2 \lambda  \nu ^2}, \ \ \ \ \ \ I_2(\lambda,\nu,3) =\frac{1}{4 \pi  \lambda +4 \pi  \nu }.
  \end{equation}
 \item  At $d=4$ we encounter the first divergence. The Laurent expansion of the formula near $d=4$ gives  
 \begin{eqnarray}
&& \left.I_2(\lambda,\nu,d)\right |_{d\approx 4} =-\frac{1}{8 \pi ^2 (d-4)}+ \frac{1-\gamma +\log (4 \pi )}{16 \pi ^2}-\frac{1}{8
   \pi ^2 (\lambda +\nu )}\cr &&+\frac{\left(1-4 \lambda ^2\right) \psi
   \left(\lambda -\frac{1}{2}\right)-\left(1-4 \nu
   ^2\right) \psi \left(\nu
   -\frac{1}{2}\right)}{32 \pi ^2 (\lambda -\nu )
   (\lambda +\nu )}
+
\ {\rm O}(d-4). \label{Ilambdanu4}
 \end{eqnarray}
\item In the limit when the two masses are equal the general formula reduces to 
\begin{equation}
I_2(\nu,\nu,d)  = \frac{ \Gamma
   \left(1-\frac{d}{2}\right) \Gamma
   \left(\frac{d-1}{2}+\nu \right) }{2^{d+1} \pi ^{\frac d 2} \nu \  \Gamma
   \left(\frac{3-d}{2}+\nu \right)} \left(\psi
   \left(\frac{3-d}{2}+\nu
   \right)-\psi \left(\frac{d-1}{2}+\nu
   \right)\right). \label{Inunu}
   \end{equation}
This expression coincides with a result of \cite{Carmi2018}. 
 
\end{enumerate}

\section{Two and three lines (bulk-to-bulk) 
\label{sec5a} and a general formula for the bubble in any background geometry}
Generalizing Eq. (\ref{bub}),  we now consider the two-line integral where the external points do not coincide:
\begin{eqnarray}
F_{\lambda\,\nu}(x,y)=
 \int_{\Lobad} G^{d}_{\lambda}(x\cdot z)  G^{d}_{\nu}(y\cdot z) \sqrt {g(z)}\, dz .  \label{diagram}
\end{eqnarray}
$F_{\lambda\,\nu}(x,y)$ solves the following equations: 
\begin{eqnarray}
&& (-\nabla^2_x+m^2_{\lambda})F_{\lambda\,\nu}(x,y)=
 \int \delta (z, x)  G^{d}_{\nu}(z, y) \sqrt g\,dz = G^{d}_{\nu}(x\cdot y), \label{011}
\\
&& (-\nabla^2_y+m^2_{\nu})F_{\lambda\,\nu}(x,y)=
 \int G^{d}_{\lambda}(x\cdot z)  \delta(y, z)  \sqrt g\,dz = G^{d}_{\lambda}(x\cdot y). \label{012}
\end{eqnarray}
Lorentz invariance implies that 
$ 
F_{\lambda\,\nu}(x,y)=F_{\lambda\,\nu}(x\cdot y)
$ 
and therefore 
\begin{eqnarray}
\nabla^2_x F_{\lambda\,\nu}(x,y)=\nabla^2_y F_{\lambda\,\nu}(x,y). 
\end{eqnarray}
Subtracting Eq. (\ref{012}) from Eq. 
 (\ref{011}) it follows that
\begin{eqnarray}
 F_{\lambda\,\nu}(x,y)= -\frac{G^{d}_{\lambda}(x\cdot y) - G^{d}_{\nu}(x\cdot y)}{m_\lambda^2-m_\nu^2}. \label{formu}
\end{eqnarray}
Let us check how this remarkable  formula works in the one-dimensional case $AdS_1$. Here 
\begin{align}& G^1_\nu(\cosh v) = \frac {e^{-\nu |v|} }{2\nu }  
\\
& F^1_{\lambda\,\nu}(a,b),=
 \int_{-\infty}^{\infty}  \frac {e^{-\lambda |v-a|} }{2\lambda } \frac {e^{-\nu |v-b|} }{2\nu } dv.
\end{align}
Suppose to fix the ideas that $a<b$. It follows that 
\begin{eqnarray}
&&F_{\lambda\,\nu}(a,b)=
 \int_{b}^{\infty}  \frac {e^{-\lambda (v-a)} }{2\lambda } \frac {e^{-\nu (v-b)} }{2\nu } dv + \int_{a}^{b}  \frac {e^{-\lambda (v-a)} }{2\lambda } \frac {e^{-\nu(b-v)} }{2\nu } dv+ \int_{b}^{a}  \frac {e^{-\lambda (a-v)} }{2\lambda } \frac {e^{-\nu (b-v)} }{2\nu } dv = \cr 
&& = \frac{1}{\lambda^2-\nu^2}\left(\frac {e^{-\nu (b-a)} }{2\nu }  - \frac {e^{-\lambda (b-a)} }{2\lambda }  \right). \label{formu10}
\end{eqnarray}
 Iterating once the above construction  we may compute the  3-line diagram with two convolutions:
\begin{eqnarray}
&&F_{\lambda\,\nu\, \kappa}(x,w)=
 \int_{\Lobad\times \Lobad} G^{d}_{\lambda}(x\cdot y)  G^{d}_{\nu}(y\cdot z)   G^{d}_{\kappa}(z\cdot w) \sqrt {g(y)}\, dy \sqrt {g(z)}\, dz  =  \cr 
&=& \frac{1}{\lambda^2-\nu^2}   \int_{\Lobad}  (G^{d}_{\nu}(x\cdot z) - G^{d}_{\lambda}(x\cdot z)) G^{d}_{\kappa}(z\cdot w) \sqrt {g(z)}\, dz  =  \cr 
&=&\frac{ G^{d}_{\kappa}(x\cdot w) }{(\kappa^2-\nu^2)(\kappa^2-\lambda^2)} +\frac{ G^{d}_{\nu}(x\cdot w) }{(\nu^2-\lambda^2)(\nu^2-\kappa^2)} +\frac{ G^{d}_{\lambda}(x\cdot w)  }{(\lambda^2-\nu^2)(\lambda^2-\kappa^2)} \label{pio}
\end{eqnarray}
a formula which is readily generalized to the $n+1$-line case.
\begin{remark}
Note that the analogous of Eq. (\ref{formu}) in flat space is made very simple by working in Fourier space. Actually,  it is evident from the above elementary derivation that Eq. (\ref{formu})  is valid for the Schwinger propagator of a Klein-Gordon field in a generic curved background ${\cal E}_d$ with Euclidean signature:
\begin{eqnarray}
&& F_{m_1 m_2}(x,y)= \int_{{\cal E}_d} G^{({\cal E})}_{m_1}(x, z)  G^{({\cal E})}_{m_2}(z, y) \sqrt {g(z)}\, dz =   \label{diagram} -\frac{G^{({\cal E})}_{m_1}(x, y) - G^{({\cal E})}_{m_2}(x, y)}{m_1^2-m_2^2}. \cr &&\label{formu2}
\end{eqnarray}
In the limiting case when the two masses coincide  this becomes
\begin{eqnarray}
&& F_{m m}(x,y)= -\frac {\partial} {\partial m^2} {G^{({\cal E})}_{m}(x, y).}
\end{eqnarray}
 Taking the limit $y\to x$ we deduce a general formula for the bubble in a generic curved background with Euclidean signature (similar remarks apply to the chronological propagator in Lorentzian signature):
 \begin{eqnarray}
bubble(m_1,m_2) = \int_{{\cal E}_d} G^{({\cal E})}_{m_1}(x, z)  G^{({\cal E})}_{m_2}(z,x) \sqrt {g(z)}\, dz= -\lim_{y\to x} \frac{G^{{(\cal E)}}_{m_1}(x, y) - G^{{(\cal E)}}_{m_2}(x, y)}{m_1^2-m_2^2}. \cr \label{formu3}
\end{eqnarray}
\end{remark}
In particular, in flat Euclidean space the propagator has the following short distance behaviour   
\begin{align}
G_m^{(M)}(r) =   
 \frac 1 {(2\pi)^{\frac d 2 }}   \left(\frac{r }{ m}\right)^{1-\frac{d}{2}}  K_{\frac{d}{2}-1}\left( m r \right)
\simeq \frac{r^{2-d}}{4 \pi ^{\frac d 2}   } 
  \Gamma \left(\frac{d}{2}-1\right)+ \frac{m^{d-2}}{(4 \pi) ^{\frac d 2}}  \Gamma \left(1-\frac{d}{2}\right); \label{sdb}
\end{align} 
here  $M$ stands for Minkowski and $r=(x_1-x_2)^2$ is the Euclidean distance. Inserting  the expansion (\ref{sdb}) in Eq. (\ref{formu2}) the diverging terms drop and the standard flat space result \cite{{Cacciatori2023a}} is recovered without any  calculation:
 \begin{eqnarray}
bubble_M(m_1,m_2)  = -\lim_{r\to 0} \frac{G^{(M)}_{m_1}(x, y) - G^{(M)}_{m_2}(x, y)}{m_1^2-m_2^2} = -\frac{\Gamma \left(1-\frac{d}{2}\right)}{(4 \pi) ^{\frac d 2} }\frac{(m_1)^{\frac d 2-1}-(m_2^2)^{\frac d 2-1}}{m_1^2-m_2^2}. \cr 
\end{eqnarray}
Similarly, in the AdS case, we may evaluate (\ref{formu2}) with  the help of   Eq. (\ref{kgtp}); here $y\to x$ means $\zeta \to 1$; at $\zeta=1$ and {\em with the condition $d<2$} that formula gives 
\begin{align}
& \lim_{\zeta\to 1} \frac{ \Gamma
   \left(\frac{d-1}{2}+\nu \right) {}_2F_1\left(\frac{d-1}{4}+\frac{\nu }{2},\frac{d+1}{4}+\frac{\nu }{2};\nu
   +1;\frac{1}{\zeta^2}\right) \,
  }{ 2\pi ^{\frac{d-1}{2}}  (2\zeta)^{ \frac{d-1}{2}+\nu }\Gamma (\nu +1)}  \  = \frac{ \Gamma \left(1-\frac{d}{2}\right) \Gamma \left(\frac{d-1}{2}+\nu\right)}{ 2^{d} \pi ^{\frac d 2}\Gamma \left(\frac{3-d}{2}+\nu \right)}.
\label{kgtp77}
\end{align}
Inserting this into Eq. (\ref{formu2}) we immediately recover  Eq. (\ref{bubads}) again without calculation:
\begin{eqnarray}
bubble _{AdS}= F_{\lambda\,\nu}(x,x) = 
 \frac{ \Gamma \left(1-\frac{d}{2}\right)}{ 2^{d} \pi ^{\frac d2}(\lambda^2-\nu^2)}\left( \frac{ \Gamma \left(\frac{d+1}{2}+\nu
   \right)}{ \Gamma \left(\frac{3-d}{2}+\nu \right)}
-  \frac{\Gamma \left(\frac{d+1}{2}+\lambda
   \right)}{ \Gamma \left(\frac{3-d}{2}+\lambda \right)} \right).
\end{eqnarray}
This formula defines  the bubble by analytic continuation as a meromorphic function of $d$.

Finally, in the de Sitter case (see Appendix \ref{aapa}) we may use Eq. (\ref{wigds}) at $\zeta= 1$, always restricting the spacetime dimension $d<2$: 
\begin{align}
& \lim_{\zeta\to -1} {\Gamma\left(\frac{d-1}2 + i \kappa\right)\Gamma\left(\frac{d-1}2 - i \kappa\right) \over
(4\pi)^{\frac d 2}\Gamma\left({d\over 2}\right)}
{}_2F_1\left(\frac{d-1}2 + i \kappa,\frac{d-1}2 - i \kappa;\ {d\over 2};\ {1-\zeta\over 2}\right ) =
\cr& =\frac{\Gamma \left(1-\frac{d}{2}\right) \Gamma \left(\frac{d-1}{2}-i \kappa \right)
   \Gamma \left(\frac{d-1}{2}+i \kappa \right)}{2^d \pi ^{\frac d 2} \Gamma \left(\frac{1}{2}-i
   \kappa \right) \Gamma \left(\frac{1}{2}+i \kappa \right)}
\end{align}
so that 
\begin{eqnarray}
bubble _{dS}=  
 \frac{ \Gamma \left(1-\frac{d}{2}\right)}{ 2^{d} \pi ^{\frac d2}(\lambda^2-\nu^2)}\left( \frac{\Gamma \left(\frac{d-1}{2}-i \nu \right)
   \Gamma \left(\frac{d-1}{2}+i \nu \right)}{\Gamma \left(\frac{1}{2}-i
   \nu \right) \Gamma \left(\frac{1}{2}+i \nu \right)}-\frac{\Gamma \left(\frac{d-1}{2}-i \lambda \right)
   \Gamma \left(\frac{d-1}{2}+i \lambda \right)}{\Gamma \left(\frac{1}{2}-i
   \lambda \right) \Gamma \left(\frac{1}{2}+i \lambda \right)} \right). \cr
\end{eqnarray}

In a completely analogous way we may fully evaluate the triangular loop  by setting $w\to x$ in Eq. (\ref{pio}); as before we do it at first assuming that $d<2$ and then generalizing the formula by analytic continuation:
 \begin{align}
&Triangle_{AdS}=
 \int_{\Lobad} G^{d}_{\lambda}(x\cdot y)  G^{d}_{\nu}(y\cdot z) G^{d}_{\kappa}(z\cdot x)  \sqrt {g(y)}\, dy  \sqrt {g(z)}\, dz  \cr 
&= \frac{ \Gamma \left(1-\frac{d}{2}\right) }{ 2^{d} \pi ^{\frac d 2}}\left(\frac{ {  \Gamma \left(\frac{d-1}{2}+\kappa\right)}{ }}{(\kappa^2-\nu^2)(\kappa^2-\lambda^2) \Gamma \left(\frac{3-d}{2}+\kappa \right)} +\frac{ {  \Gamma \left(\frac{d-1}{2}+\nu\right)}{ }}{(\nu^2-\kappa^2)(\nu^2-\lambda^2) \Gamma \left(\frac{3-d}{2}+\nu \right)} \right. \cr &\left.+\frac{   \Gamma \left(\frac{d-1}{2}+\lambda\right)}{(\lambda^2-\kappa^2)(\lambda^2-\nu^2) \Gamma \left(\frac{3-d}{2}+\lambda \right)} \right).
\end{align}
Finally, it is worth to underline a remarkable property of Eq. (\ref{formu}): it can be used also when the Schwinger function is known only approximately. A simple example is an application to Quantum Field Theory on the Rindler space. We refer to \cite{Svaiter,GL1,GL2,Linet}. Beyond the Minkowski vacuum $|M\rangle$ one can consider the Fulling-Rindler vacuum $|R\rangle$. Let us call $G_A^{\beta,d}(x,y;m)$ the Schwinger function at inverse temperature $\beta$ in $d$ spacetime dimensions and for the vacuum state $|A\rangle$, for a scalar field of mass $m$. In coordinates $(\tau,\xi,x^1,\ldots,x^{d-2})$ such that the Euclidean Rindler metric is
\begin{align}
    ds^2=\xi^2 d\tau^2+d\xi^2+\delta_{ij}dx^i dx^j,
\end{align}
it is manifest that the Euclidean Rindler's wedge has the topology of a cylinder with period normalized to $2\pi$ for the Euclidean time. As it is well known, this leads to the identity
\begin{align}
    G_M^{\infty,d}(x,y;m)=G_R^{2\pi,d}(x,y;m),
\end{align}
which corresponds to the fact that the zero temperature Minkowski vacuum appears as a thermal state at temperature $\frac 1{2\pi}$ to the Rindler vacuum (see \cite{Svaiter}, and reference therein). It is interesting to consider also the Schwinger function $G_R^{\infty,4}(x,y)$. Assuming $|\tau_x-\tau_y|<\pi$, one can prove (see \cite{Linet}) that 
\begin{align}
    G_R^{\infty,4}(x,y;m)=& G_R^{2\pi,4}(x,y;m)+\frac m{8\pi^3} \int_0^\infty du \frac {K_1(mR(u))}{R(u)}F_\infty(u,\tau_x-\tau_y), \\
    R(u)^2=& \xi_x^2+\xi_y^2+2\xi_x\xi_y \cosh u+(x_1-y_1)^2+(x_2-y_2)^2, \\
    F_\infty(u,v)=& -2 \frac {v+\pi}{(v+\pi)^2+u^2}+2 \frac {v-\pi}{(v-\pi)^2+u^2}.
\end{align}
In this way, the UV divergences are confined in $G_R^{2\pi,4}(x,y;m)=G_M^{\infty,4}(x,y;m)$. At this stage, an immediate application of the above general strategy shows that
\begin{align}
   bubble_{Rindler}(m_1,m_2) =bubble_M(m_1,m_2).
\end{align}
Another possible application, which we don't include here, could be to compute the bubble integral (and the effective potential), for a massive scalar field in the far region of a Schwarzschild black hole of mass $M$ (t.i., for $M/R$), at the first order in $M$, starting from the results in \cite{Schwarz}.

\section{A remark about bulk diagrams versus Witten diagrams}
\label{6}


 Now we want to consider the limit of the expression (\ref{formu}) when $x$ and/or $y$ tend to the boundary. To this aim it is useful to give coordinates to the Lobachevsky space as follows:
 \begin{align}
& x^0= \cosh v + \frac 12 e^v r^2 , \ \ x^i = e^v r^i,  \ \ x^{d-1}  = \sinh v - \frac 12 e^v r^2 \\
& y^0= \cosh v' + \frac 12 e^{v '}{r'}^2 , \ \ y^i = e^{v '}{r'},  \ \ x^{d-1}  = \sinh v' - \frac 12  e^{v '}{r'}^2
 \end{align}
 so that 
 \begin{align}
x\cdot y = \cosh (v-v') + \frac 12 e^{v+v'} (r-r')^2 .
 \end{align}
 Let $v$ tend to plus infinity while $v'$ is held fixed: 
  \begin{align}
x\cdot y \simeq \frac {e^{v}}2 (e^{-v'}  +  e^{v'} (r-r')^2 ) = e^v (\xi \cdot y);
 \end{align} 
 in the second equality  we have introduced the lightlike vector
 \begin{align}
& \xi^0= \frac {1+r^2}2  , \ \ \xi^i =  r^i,  \ \ \xi^{d-1}  = \frac {1-r^2}2.
 \end{align}
 Eq. (\ref{kgtp}) provides the behaviour of the propagator  for large values of $\zeta$; at leading order 
\begin{align}
&w^d_{\nu}(\zeta) \simeq \frac{ \Gamma
   \left(\frac{d-1}{2}+\nu \right)  \,
  }{ 2\pi ^{\frac{d-1}{2}}  (2\zeta)^{ \frac{d-1}{2}+\nu }\Gamma (\nu +1)}. \label{kgtpapprox}
\end{align}
 It follows that
 \begin{eqnarray}
&& F_{\lambda\,\nu}(x,y)\simeq \frac{G^{d}_{\nu}(e^v (\xi \cdot y)) - G^{d}_{\lambda}(e^v (\xi \cdot y))}{\lambda^2-\nu^2} \simeq \cr && \simeq \frac{1}{\lambda^2-\nu^2}
\left(\frac{ 
  e^{ -\left( \frac{d-1}{2}+\nu\right) v } \Gamma\left( \frac{d-1}{2}+\nu \right)  \,
  }{ 2\pi ^{\frac{d-1}{2}}  (2 \xi \cdot y)^{ \frac{d-1}{2}+\nu }\Gamma (\nu +1)} -\frac{ 
  e^{ -\left( \frac{d-1}{2}+\lambda\right) v } \Gamma\left( \frac{d-1}{2}+\lambda \right)  \,
  }{ 2\pi ^{\frac{d-1}{2}}  (2 \xi \cdot y)^{ \frac{d-1}{2}+\lambda }\Gamma (\lambda +1)} \right).
\end{eqnarray}
To extract a finite limit we now need to specify something about the masses. Let us suppose that $\lambda>\nu$; in this case standard procedure \cite{witten-adscft, bemb}, using the we get that
 \begin{eqnarray}
&& e^{ \left( \frac{d-1}{2}+\nu\right) v }F_{\lambda\,\nu}(x,y)  \to F_{\lambda\,\nu}(\xi,y) =\frac{1}{\lambda^2-\nu^2}
\left(\frac{ 
 \Gamma\left( \frac{d-1}{2}+\nu \right)  \,
  }{ 2\pi ^{\frac{d-1}{2}}  (2 \xi \cdot y)^{ \frac{d-1}{2}+\nu }\Gamma (\nu +1)}  \right).
\end{eqnarray}
 Let now $v'$ tend to plus infinity: 
  \begin{align}
\xi \cdot y = e^{v'} (\xi \cdot \eta);
 \end{align} 
where we have introduced the lightlike vector
 \begin{align}
& \eta^0= \frac {1+{r'}^2}2  , \ \ \xi^i =  {r'}^i,  \ \ \xi^{d-1}  = \frac {1-{r'}^2}2.
 \end{align}
 Proceeding as before we get ($\lambda>\nu$)
  \begin{eqnarray}
&& e^{ \left( \frac{d-1}{2}+\nu\right) v' }F_{\lambda\,\nu}(\xi ,y)  \to F_{\lambda\,\nu}(\xi,\eta) =\frac{1}{\lambda^2-\nu^2}
\left(\frac{ 
 \Gamma\left( \frac{d-1}{2}+\nu \right)  \,
  }{ 2\pi ^{\frac{d-1}{2}}  (2 \xi \cdot \eta)^{ \frac{d-1}{2}+\nu }\Gamma (\nu +1)}  \right) \label{limit}
\end{eqnarray}
 Had we performed the limit in the opposite order or else  had we put  $v=v'$ and taken the limit together, the situation would not have  changed: when $\lambda>\nu$ the finite  result extracted from Eq. (\ref{formu}) is  (\ref{limit}).

 Let us now revert the order of the operations and take the dominant terms in Eq. (\ref{diagram}) before integration. We are led to the following Witten diagram which is the simplest among the ones considered in \cite{Pene}:
  \begin{eqnarray}
\tilde F_{\lambda\,\nu}(\xi ,\eta)=
 \int_{\Lobad} \frac{ \Gamma
   \left(\frac{d-1}{2}+\lambda \right)  \,
  }{ 2\pi ^{\frac{d-1}{2}}  (2(\xi\cdot z) )^{ \frac{d-1}{2}+\lambda }\Gamma (\lambda +1)} \frac{ \Gamma
   \left(\frac{d-1}{2}+\nu \right)  \,
  }{ 2\pi ^{\frac{d-1}{2}}  (2(\eta \cdot z))^{ \frac{d-1}{2}+\nu }\Gamma (\nu +1)} \sqrt {g(z)}\, dz  \cr 
\end{eqnarray}
We see that this Witten diagram cannot be finite. Otherwise it should be function of $(\xi\cdot\eta)$ homogeneous of degree $-\frac{d-1}{2}-\lambda$ and, at same time, of degree $-\frac{d-1}{2}-\nu$, a thing which is logically impossible. This is also what happens in \cite{Pene}, eq.(24), where for $n=2$ the integral is proportional to 
\begin{align}
    \int_0^\infty \frac {dt_1}{t_1} t_1^{\Delta_1} \int_0^\infty \frac {dt_2}{t_2} t_2^{\Delta_2} e^{-t_1t_2 (x_1-x_2)^2}=
    \Gamma(\Delta_2) ((x_1-x_2)^2)^{2\Delta_2}\int_0^\infty \frac {dt_1}{t_1} t_1^{\Delta_1-\Delta_2},
\end{align}
which is divergent.

We are led to the conclusion that, at least in this example, this Witten diagram does not arise as the limit of  corresponding diagrams where the endpoints lie in the bulk. This example  may however be pathological.

In the three-line case the Witten diagram is completely determined by Lorentz invariance and by the homogeneity in the conical variables $\xi_1$, $\xi_2$ and $\xi_3$; its  simple formula that the structure of a star-triangle relation \cite{us}:
  \begin{eqnarray}
&&\widetilde G_{\nu_1\nu_2\nu_3}(\xi_1 ,\xi_2,\xi_3)= c_0(\nu_1,\nu_2,\nu_3)
 \int_{\Lobad} \frac{ 1}{(\xi_1\cdot z) ^{ \frac{d-1}{2}+\nu_1 }(\xi_2\cdot z) ^{ \frac{d-1}{2}+\nu_2 }(\xi_3\cdot z) ^{ \frac{d-1}{2}+\nu_2 }} \sqrt {g(z)}\, dz  \cr &&    \cr  &&= c(\nu_1,\nu_2,\nu_3)(\xi_1\cdot \xi_2)^{\frac{1-d}{4} +\frac {\nu_3}2-\frac{\nu_1}2-\frac{\nu_2}{2}}(\xi_2\cdot \xi_3)^{\frac{1-d}{4} +\frac {\nu_1}2-\frac{\nu_2}2-\frac{\nu_3}{2}}(\xi_1\cdot \xi_3)^{\frac{1-d}{4} +\frac {\nu_2}2-\frac{\nu_1}2-\frac{\nu_3}{2}} \label{w3}
\end{eqnarray}
where
\begin{align}
&c(\nu_1,\nu_2,\nu_3)   =  \cr 
&= \frac{  \Gamma
   \left(\frac{d-1+2 \nu_1+2 \nu_2-2 \nu_3}{4} \right) \Gamma
   \left(\frac{d-1+2 \nu_1-2 \nu_2+2 \nu_3}{4} \right) \Gamma
   \left(\frac{d-1-2 \nu_1+2 \nu_2+2 \nu_3}{4} \right) \Gamma
   \left(\frac{d-1+2 \nu_1+2 \nu_2+2 \nu_3}{4} \right)}{ 2^{\frac{3 d+2 \nu_1+2 \nu_2+2 \nu_3+13}{4} } \pi ^{d-1}\Gamma
   (\nu_1+1) \Gamma (\nu_2+1) \Gamma (\nu_3+1)}.\cr & 
\end{align}
This also coincides with formula (24) in \cite{Pene} once expanded by using the formula
\begin{align}
   \int_0^\infty \frac {dt_1}{t_1} t_1^{\Delta_1} \int_0^\infty \frac {dt_2}{t_2} t_2^{\Delta_2} \int_0^\infty \frac {dt_3}{t_3} t_3^{\Delta_3} 
   e^{-t_1t_2 a_3^2-t_1t_3 a_2^2 -t_2t_3 a_1^2}=\frac 12 \prod_{j=1}^3 \Gamma(-\sigma_j) a_j^{2\sigma_j}
\end{align}
with $\sigma_j=\Delta_j-\frac {\Delta_1+\Delta_2+\Delta_3}2$.

The corresponding 3-line bulk diagram is given by
 \begin{eqnarray}
&&G_{\lambda\,\nu\,\kappa}(x,y,w)=
 \int_{\Lobad} G^{d}_{\lambda}(x\cdot z)  G^{d}_{\nu}(y\cdot z) G^{d}_{\kappa}(w\cdot z)  \sqrt {g(z)}\, dz = {\bf G}_{\lambda\,\nu\,\kappa}(a,b,c). \label{diagram3} \\
&& a= x\cdot y,\ \ b= y\cdot w, \ \ c=w\cdot x 
\end{eqnarray}
At the moment we are unable to determine whether, contrary to what happens in the two-line case,  the Witten diagram (\ref{w3}) may be obtained as  boundary limit of (\ref{diagram3}) or not.

 \newpage


\newpage


\section{The K\"all\'en-Lehmann expansion of the bubble with two independent masses}
\label{kal}
In this section we come back and elaborate on the AdS K\"all\'en-Lehmann expansion formula which we found and proved in \cite{begmp2} and that appeared also in a later paper \cite{fitzpatrick}.

Recall that for  any given analytic  two-point function $W(z_1,z_2)$  of a local quantum field in Minkowski space  satisfying the Wightman axioms \cite{Streater} there exists a tempered weight $\rho$ with support in the positive real axis such that
\beq 
W(z_1, z_2) = \int_0^\infty \rho (m^2)  W_m(z_1, z_2) dm^2;
\endq
 here 
 \beq 
W_m(z_1, z_2) = \frac 1{(2\pi)^d}\int e^{-ik(z_1-z_2) } \theta(k^0) \delta(k^2-m^2) dk
\endq
 is the analytic continuation of the two-point function of a Klein-Gordon field with mass $m$; $z_1$ belongs to the past tube and $z_2$ to the forward tube if the complex Minkowski space \cite{Streater}.
The K\"all\'en-Lehmann weight $\rho (m^2)$ is a positive measure if and only if the boundary value $W(x_1,x_2)$ is a distribution of positive type. 

In particular, for any two given
masses $m_1$ and $m_2$ the weight for the bubble 
\beq 
W_{m_1}(z_1, z_2) W_{m_2}(z_1, z_2) = \int_{(m_1+m_2)^2}^\infty \rho (m^2:m_1,m_2)  W_m(z_1, z_2) dm^2
\endq 
is easily obtained by Fourier transform. {The above relation can be extended to Schwinger functions at non-coincident Euclidean points. A distributional extension to coincident Euclidean points is also possible if no renormalization is necessary,
e.g. for low dimension.}

As regards the general case of AdS fields on the covering  manifold $\wt{AdS_d}$,  
it is of course possible to  consider a  general superposition  {\em \`a la}  K\"all\'en-Lehmann  as follows:
\beq 
w(\zeta) = \int_{-1}^\infty \rho_1 (\n)  w^d_{\nu}(\zeta) d\nu+\int_{-1}^{1} \rho_2 (\n)  w^d_{-\nu}(\zeta) d\nu.
\endq
{We expressed the above relation using the reduced maximally analytic two-point function; under suitable condition this implies the same representation for the Euclidean propagator:}
\beq 
G(u) = \int_{-1}^\infty \rho_1 (\n)  G^d_{\nu}(u) d\nu+\int_{-1}^{1} \rho_2 (\n)  G^d_{-\nu}(u) d\nu.
\endq
However, when we restrict our attention to the bubble, we do not expect an integral as in flat space; even though we consider fields with non integer masses that are well-defined only  on the covering manifold, their correlation functions are  quasi-periodic. {\em This calls for a K\"all\'en-Lehmann series: the weight has to be a series of pure point measures.} 

Before addressing the general case, discussing  two concrete examples will clarify the point.

\subsection{An almost trivial example: $\wt{AdS}_3$}
In odd spacetime dimension the Schwinger function can be written in terms of elementary functions.  To this aim, it is useful to parametrize the invariant variable by an hyperbolic angle: $u= \cosh v$. 
In $d=3$ this  gives the following expressions: 
\cite[Eq. 12, p 150]{bateman}:
\begin{align}
&  Q_{\n -{1\over 2}}^{1\over 2}(\cosh v) =  e^{{i\pi \over 2}-\nu v} \sqrt {\frac \pi {2\sinh v}}, \label{QQ}
\\ &
G^3_\n(\cosh v) = {e^{-{i\pi \over 2}} Q_{\n-{1\over 2}}^{1\over 2}(\cosh v)  \over 2\pi \sqrt{2 \pi \sinh v}} 
= \frac {e^{-\nu v}}{4 \pi \sinh  v}  .
\end{align}
The bubble with two independent  mass parameters $\lambda$ and $\nu$ is readily decomposed into its K\"all\'en-Lehmann series  by an elementary manipulation:
\begin{align}
G^3_\lambda(\cosh v) G^3_\nu(\cosh v)& = \frac {e^{-(\lambda+\nu) v}}{16 \pi^2 \sinh^2  v} = \frac {e^{-(\lambda+\nu+1) v}}{8 \pi^2 \sinh  v(1-e^{-2v}) } = \sum_{k=0}^\infty \frac {e^{-(\lambda+\nu+1+2k ) v}}{8 \pi^2 \sinh  v }  = \cr &= \frac{1}{2\pi} \sum^\infty_{k=0} G^3_{2k+1+\lambda+\nu}(\cosh v). 
\end{align}
So, we get a K\"all\'en-Lehmann series as opposed to an integral, with a weight that in this simple case is just a constant.

\subsection{A less elementary example: $\wt{AdS}_5$}
 In view of its importance in the AdS/CFT correspondence, let us discuss now the  less trivial five-dimensional case: here 
\begin{align}
G^5_\nu(\cosh v)  ={1 \over (2\pi)^{5\over 2}} (\sinh v)^{-\frac{3}2}e^{-{3 \pi i\over 2}} Q_{\n-{1\over 2}}^{3\over 2}(\cosh v) = 
{ (\nu+1) e^{-(\nu-1)v}-(\nu-1) e^{-(\nu+1)v}  \over 16\pi^2 \sinh^3 v}; \label{w5}
\end{align}
to derive the above formula we used  Eq. (\ref{QQ}) and the recurrence relation \cite[Eq. 4, p 160]{bateman} 
\begin{equation}
Q^{\mu+1}_{\nnu}(z) = 
\frac{(z^2-1)^{-\frac 12 }}{(\nnu-\mu+1)}  \left((\nnu-\mu){(2\nnu + 1)} Q^{\mu}_{\nnu+1}(z) - {(\nnu+\mu)(\nnu+\mu+1)})Q^{\mu}_{\nnu-1}(z)\right). \end{equation}
Now, taking into account the identity
\begin{eqnarray}
\frac 1 {4\sinh^3 v} = 
\sum_{k=0}^{\infty} {k(k-1)}e^{-(2k-1)v} ,\ \ \ \ \Re\, v>0
\end{eqnarray}
we get a  series expansion which is not yet a K\"all\'en-Lehmann expansion:
\begin{align}
 G^5_\lambda(\cosh v)  
G^5_\nu(\cosh v)  &= \sum_{k=0}^{\infty}{ (\lambda +1) (\nu +1)+2 k^2+2 k (\lambda +\nu +1)
\over 8\pi^4 \sinh^3 v} e^{-(2k+\lambda+\nu+1)v}.
\end{align}
For  $\lambda, \  \nu$ and $\lambda+\nu$ greater than $-1$ all the  terms at the r.h.s. are positive. We may therefore rearrange the series as it is absolutely convergent. 

The construction goes  as follows: in the first term  of the series ($k=0$) we replace  the exponential $e^{-(\lambda+\nu+1)v}$ with the Schwinger function $G^5_{\lambda+\nu +2}(\cosh v)$ and subtract the corresponding term proportional to $e^{-(\lambda+\nu+3)v}$ from the second term $(k=1)$. Then we iterate the procedure. It results  the  following K\"all\'en-Lehmann series  for the bubble in $\wt {AdS_5}$: 
\begin{align}
& G^5_\lambda(\cosh v)  
G^5_\nu(\cosh v) = \cr & = 
\sum_{k=0}^\infty  {(k+1)(\lambda +k+1) (\nu +k+1)(\lambda+\nu+k+1) 
\over 2\pi^2 (\lambda +\nu +2k+1)(\lambda +\nu +2k+3)} G^5_{\lambda+\nu+2k +2}(\cosh v).
\end{align}

Now let us consider the general problem. The following two theorems were proved in \cite{begmp2} and solve the general problem of finding the weight for the K\"all\'en-Lehmann series  of the bubble in $\wt {AdS_d}$. The first theorem concern the ultraspherical Gegenbauer functions; it holds for integer values of the parameters:

\begin{theorem}
[\cite{begmp2}]
Let  $c_\delta(m,n|l)$ be given by 
\begin{align}
c_\delta(m,n|l) 
=\frac{\pi {\alpha_\delta\left(\frac{l+m-n}{2}\right)}
 {\alpha_\delta\left(\frac{l-m+n}{2}\right)}
 {\alpha_\delta\left(\frac{l+m+n}{2}+\delta\right)}
{\alpha_\delta\left(\frac{l-m-n}{2}-\delta\right)}}{ \alpha_\delta(l)
\alpha_\delta\left(l+\delta \right)} \ .
\label{x.3}\end{align}
where
\beq
\alpha_\delta(s)
 =\ \ \frac{\Gamma(s+\delta)}{\Gamma(\delta)\Gamma(s+1)}, \ \
\label{s.16}
\endq
(i) Let $m$ and $n$ be non-negative integers and suppose that one of
the two following conditions is satisfied 

\noindent (a) $\delta = \half -r$, $r \ge 1$ is an integer
such that $m \ge 2r$ and $n \ge 2r$;

\noindent (b) $2\delta$ is a strictly positive integer; \\ [10pt] 
then 
\beq
D^{\delta}_{m}(z) D^{\delta}_{n}(z) =
\sum_{\begin{array}{c}
\scriptstyle l = m+n+2\delta + 2k\cr
\scriptstyle k \in \bZ,\ \  k\ge 0
\end{array}}
c_\delta(m,n|l) D^{\delta}_{l}(z)
\label{x.13}\endq
holds with uniform convergence for $z$ in any
compact subset of $\bC \setminus [-1,\ 1]$.\\
(ii) If $m$, $n$ and $2\delta$ are not necessarily integers but the
following conditions are satisfied:
\beq
\delta >0,\ \  m+2\delta >0,\ \  n+2\delta >0,\ \  m+\delta+1 >0,\ \
n+\delta+1 >0,\ \  m+n+2\delta+1 \ge 0\ ,
\label{x.12}\endq 
then (\ref{x.13}) holds  with uniform convergence for $z$ in any
compact subset of $\bC \setminus (-\infty,\ 1]$,
\end{theorem}
The proof of this theorem can be  found in \cite{begmp2}; it actually holds
in the sense of functions with tempered behavior in $\wt \Delta_1$, so that the conclusion also holds for the boundary values of both
sides in (\ref{x.13}).

{
\begin{remark}\em
\label{formalseries}
If in Eq. (\ref{x.13}) both sides are multiplied by $z^{m+n+4\delta}$,
both sides become formal power series in $z^{-2}$ (with coefficients
meromorphic in the parameters). It is proved in \cite{begmp2} that
these two formal power series are equal without any restriction on
the parameters. Moreover the series converges for $|z^2| >1$, as the
l.h.s. is holomorphic there. This does not prove that the series (in $k$)
on the r.h.s. converges: the theorem gives sufficient conditions for that.
\end{remark}
}

\vskip 10 pt
The second theorem gives the K\"all\'en-Lehmann expansion for AdS quantum fields on the covering manifold with two distinct mass parameters. It was shown under rather restricitive conditions but it holds for continuous values of the real mass parameters.
\begin{theorem}[\cite{begmp2}]
\label{k-l}
Let $\bu = {d-1\over 2}$. Under the conditions \beq
\bu >0,\ \  \lambda +\bu >0,  \ \ \nu  +\bu >0,\ \  \lambda+1 >0,\ \  \nu +1 >0,\ \
\lambda+\nu +1 \geq 0,\ \  
\label{x.14}\endq
\beq
w^d_{\lambda}(\zeta)\,w^d_{\nu}(\zeta) =\sum_{k=0}^\infty \rho_d(k;\lambda,\nu)
\,w^d_{\bu +2k  +\lambda+\nu}(\zeta)\ , \ \ \zeta \in \wt\Delta_1
\label{x.15}\endq
where the weight is given by 
\begin{align}
\rho_d(k;\lambda,\nu)& = {(2\bu+4k+2\lambda+2\nu)\Gamma(1+2k+\lambda+\nu)\over
4\pi^\bu\Gamma(\bu)\Gamma(2\bu+2k+\lambda+\nu)} \times \cr
&\times {\Gamma(\bu+k)\Gamma(\bu+\lambda+k)\Gamma(\bu+\nu+k)
\Gamma(\bu+\lambda+\nu+k)\over
\Gamma(1+k)\Gamma(1+\lambda+k)\Gamma(1+\nu+k)
\Gamma(1+\lambda+\nu+k)}\ .
\label{kl}\end{align}

\end{theorem}

\begin{remark}  {\em  As anticipated,  for any given two masses $\lambda$ and $\nu$ the KL expansion of the bubble  $w^d_{\lambda}(\zeta)\,w^d_{\nu}(\zeta)$ is a series and not an integral. This is a leftover of the periodicity of the true AdS manifold.

In fact
the statement of the theorem holds under much weaker conditions as we explicitly show in the following examples. Theorem \ref{k-l4} below gives weaker sufficient
conditions
for the validity of the statement of Theorem \ref{k-l}}
\end{remark}
\subsection{Cases of $\wt{AdS}_1$ and $\wt{AdS}_{-1}$ }

The proof of the above formula is valid only for $d>1$. This does not mean that there is no K\"all\'en-Lehmann  expansion for $d\leq 1$. Let us consider indeed the simplest case $d=1$; the Schwinger function now is really elementary:
\begin{align}G^1_\nu(\cosh v) = \frac {e^{-\nu |v|} }{2\nu }  
\end{align}
and the K\"all\'en-Lehmann series has only one nonzero term:
\begin{align} 
G^1_\lambda(\cosh v)G^1_\nu(\cosh v) & = \frac {e^{-(\lambda+\nu) |v|} }{4\lambda \nu }  = \frac{\lambda+ \nu } {2\lambda\nu}G^1_{\lambda+\nu}(\cosh v)
\end{align}
This is perfectly coherent with the above formula (\ref{kl}); 
indeed, only the first coefficient has a non zero limit at $d=1$:
\beq
\lim_{d\to 1} \rho_d(0;\lambda,\nu) = \frac{\lambda+ \nu } {2\lambda\nu},\ \ \ \ \lim_{d\to 1} \rho_d(k\not= 0 ;\lambda,\nu) = 0.
\endq
The above feature is shared by  every odd negative spacetime dimension. Let us consider as a further example $\wt{AdS}_{-1}$ : the general formula for the propagator (\ref{kgtps}) here gives
\begin{align}
&G^{-1}_\nu(\cosh v) =
{(2\pi)^{1\over 2}} (\sinh v)^{\frac{3}2}
e^{i\pi{3\over 2}} Q_{-{1\over 2}+\n}^{-{3\over 2}}(\cosh v).
\end{align}
Taking into account Eq. (\ref{w5}) and the relation \cite[3.2 (2) p. 140]{bateman}
\beq
e^{i\pi\mu}\Gamma(\nnu+\mu+1)Q_\nnu^{-\mu}(z) =
e^{-i\pi\mu}\Gamma(\nnu-\mu+1)Q_\nnu^\mu(z)\ 
\label{wq.40}\endq
we get 
\begin{align}
&G^{-1}_\nu(\cosh v) =\frac \pi2  \frac{\ \Gamma (\nu -1)}{ \Gamma (\nu +2)}  \left((\nu +1) e^{-(\nu -1) v}-(\nu -1) e^{-(\nu +1)
   v}\right).
\end{align}
A direct verification shows that the K\"all\'en-Lehmann series here consists only of two terms
\begin{align}
G^{-1}_\lambda(\cosh v)G^{-1}_\nu(\cosh v)& =
\frac{\pi  (\lambda +\nu -2) (\lambda +\nu -1)}{2\lambda  \nu  (\lambda -1)  (\nu -1)}G^{-1}_{\lambda+\nu-1}(\cosh v)+ \cr & -\frac{\pi  (\lambda +\nu +1) (\lambda +\nu +2)}{2 \lambda \nu  (\lambda +1)  (\nu +1)} G^{-1}_{\lambda+\nu+1}(\cosh v) 
\end{align}
in accordance with our general result (\ref{kl}). 
\vskip 10 pt

\begin{theorem}
\label{k-l4}
Let $\bu = {d-1\over 2}$. Let $V$ be the open complex set\footnote{
Recall that $\bZ_-$ is the set of non-positive integers and
$\bZ_+$ is the set of non-negative integers.}:
\beq
V = \{(\bu,\ \lambda,\ \nu)\in \bC^3\ :\ 
\bu+\lambda \notin \bZ_-\,,\ \ \bu+\nu \notin \bZ_-\,,\ \ 
\bu+\lambda +\nu \notin \bZ_-\,,\ \ \lambda +\nu\notin \bZ_- \}
\label{zm.10}\endq
For $(\bu,\ \lambda,\ \nu)\in V$, $\z\in \wh \Delta_1$ and any integer $k\ge 0$
let
\beq
S_k(\bu,\ \lambda,\ \nu,\ \z) = \rho_d(k;\lambda,\nu)\,
\z^{2\bu+\lambda+\nu}w^d_{\bu  +\lambda+\nu+2k}(\zeta)\ , 
\label{zm.20}\endq
where $\rho_d(k;\lambda,\nu)$ is given by (\ref{kl}). 
Then for all $(\bu,\ \lambda,\ \nu)\in V$ and $\z\in \wh \Delta_1$,
\beq
\z^{2\bu+\lambda+\nu}w^d_{\lambda}(\zeta)\,w^d_{\nu}(\zeta) =
\sum_{k=0}^\infty S_k(\bu,\ \lambda,\ \nu,\ \z)\ .
\label{zm.40}\endq
Here the series on the rhs converges absolutely and uniformly
on every compact subset of $V\times \wh \Delta_1$.
Moreover (\ref{zm.40}) also holds for the boundary values of both
sides from $\ZZ_{1-}\times \ZZ_{1+}$ in the sense of tempered 
distributions.

\end{theorem}
Recall that (see (\ref{kgtp})), for any integer $p \ge 0$,
$\z \mapsto \z^{\bu+\nu-2p} w_\nu^d(\z)$ is even and holomorphic in $\wh\Delta_1$. This makes it possible to express the result 
(\ref{zm.40}) as an equality of functions analytic in $\wh \Delta_1$.

The proof of this theorem is given in Appendix \ref{apkl4}.

\section{The Tadpole and the Sunset} \label{8}
The tadpole (in a trilinear interaction or cubic self-interaction) is readily evaluated  using the K\"all\'en-Lehmann expansion (\ref{x.15}):
\begin{eqnarray}
F^{}_{\lambda,\,\n \sigma}(x)=
 \int_{\Lobad} G^{d}_{\lambda}(x\cdot z) G^{d}_{\n}(z\cdot w)G^{d}_{\sigma}(z\cdot w)\sqrt {g(w)}\, dw\sqrt {g(z)}\, dz = \cr 
= \sum_{k=0}^\infty\rho(k;\n,\sigma)
 \int_{\Lobad} G^{d}_{\lambda}(x\cdot z) G_{\bu +2k  +\n+\sigma}(z\cdot w)\sqrt {g(w)}\, dw\sqrt {g(z)}\, dz.   \label{diagram34}
\end{eqnarray}
By using  Eqs. (\ref{I1}) and (\ref{formu}) this becomes
\begin{align}
F^{}_{\lambda,\,\n \sigma}(x)=&
  -\sum_{k=0}^\infty \frac {\rho(k;\kappa,\sigma)}{\lambda^2-(\bu+2k+\n+\sigma)^2}
 \int_{\Lobad} \Big(G^{d}_{\lambda}(x\cdot w)- G_{\bu +2k  +\n+\sigma}(x\cdot w) \Big) \sqrt {g(w)}\, dw\cr
 =& \frac {f(\lambda,\kappa,\sigma)}{\bu^2-\lambda^2} +  g(\lambda,\kappa,\sigma),
\end{align}
where 
\begin{align}
   &f(\lambda,\kappa,\sigma) =  \sum_{k=0}^\infty 
   \left(\frac{\rho(k;\kappa,\sigma) }{2 \lambda  (2 k+\delta +\kappa +\lambda +\sigma )}-\frac{\rho(k;\kappa,\sigma) }{2 \lambda  (2 k+\delta +\kappa -\lambda +\sigma )}\right) \\
  & g(\lambda,\kappa,\sigma)  =  \sum_{k=0}^\infty 
   \left(-\frac{\rho(k;\kappa,\sigma) }{2 \delta  (\delta^2 -\lambda^2 )  (2 k+\kappa +\sigma )}+\frac{\rho(k;\kappa,\sigma) }{2 \delta   (\delta^2 -\lambda^2 )   (2 k+2\delta+\kappa +\sigma )} \right.\cr   &+\left.\frac{\rho(k;\kappa,\sigma) }{2 \lambda  (\delta^2 -\lambda^2 )   (2 k+\delta +\kappa
   -\lambda +\sigma )}-\frac{\rho(k;\kappa,\sigma) }{2 \lambda   (\delta^2 -\lambda^2 )(2 k+\delta +\kappa
   +\lambda +\sigma )}\right)
\end{align}
Defining 
\begin{align}
&F_{tad}(x) =  \sum_{k=0}^\infty \frac{  \rho(k;\kappa,\sigma) }{2k +x} = \cr &=
 \frac{\pi ^{-\delta } \delta  (2+\kappa +\sigma ) \Gamma (1+\delta +\kappa ) \Gamma
   (1+\delta +\sigma ) \Gamma (1+\delta +\kappa +\sigma ) \,
}{(2+x) \Gamma (2+\kappa ) \Gamma (2+\sigma
   ) \Gamma (2+2 \delta +\kappa +\sigma )} \times \cr &\times {}   _7F_6\left(\begin{array}{l}
   1+\frac{x}{2},1+\delta ,1+\delta +\kappa ,\frac{3}{2}+\frac{\kappa
   }{2}+\frac{\sigma }{2},2+\frac{\kappa }{2}+\frac{\sigma }{2},1+\delta +\sigma
   ,1+\delta +\kappa +\sigma \cr 2+\frac{x}{2},2+\kappa ,1+\delta +\frac{\kappa
   }{2}+\frac{\sigma }{2},\frac{3}{2}+\delta +\frac{\kappa }{2}+\frac{\sigma
   }{2},2+\sigma ,2+\kappa +\sigma \end{array};1\right) +\cr &+
   \frac{\pi ^{-\delta } \Gamma (\delta +\kappa ) \Gamma (\delta +\sigma ) \Gamma (1+\delta
   +\kappa +\sigma ) \,}{2 x \Gamma (1+\kappa )
   \Gamma (1+\sigma ) \Gamma (2 \delta +\kappa +\sigma )} \cr & \times {} \ _7F_6\left(\begin{array}{l}\frac{x}{2},\delta ,\delta +\kappa
   ,\frac{1}{2}+\frac{\kappa }{2}+\frac{\sigma }{2},1+\frac{\kappa }{2}+\frac{\sigma
   }{2},\delta +\sigma ,\delta +\kappa +\sigma \cr 1+\frac{x}{2},1+\kappa ,\delta
   +\frac{\kappa }{2}+\frac{\sigma }{2},\frac{1}{2}+\delta +\frac{\kappa
   }{2}+\frac{\sigma }{2},1+\sigma ,1+\kappa +\sigma\end{array} ;1\right)
\end{align}
we get 
\begin{align}
& F^{}_{\lambda,\,\n \sigma}(x)=   \frac {F_{tad}(2\delta +\kappa  +\sigma)-F_{tad}(\kappa +\sigma)}{2\delta (\bu^2-\lambda^2)}
\end{align}
which corresponds to (\ref{I1}) with a multiplicative renormalization of the mass.

We may also  compute in a similar way the general one-loop correction to the (bulk-to-bulk) diagram (\ref{diagram}) for arbitrary masses in the loop:
\begin{eqnarray}
F^{(1)}_{\lambda\,\nu}(x,y)=
 \int_{\Lobad} G^{d}_{\lambda}(x\cdot z) G^{d}_{\kappa}(z\cdot w)G^{d}_{\sigma}(z\cdot w) G^{d}_{\nu}(w\cdot y)\sqrt {g(w)}\, dw\sqrt {g(z)}\, dz .  \label{diagram33}
\end{eqnarray}
First, we linearize (\ref{diagram33}) using the K\"all\'en-Lehmann expansion (\ref{x.15}):
\begin{eqnarray}
F^{(1)}_{\lambda\,\nu}(x,y)= \sum_{k=0}^\infty\rho(k;\kappa,\sigma)
 \int_{\Lobad} G^{d}_{\lambda}(x\cdot z) G_{\bu +2k  +\n+\sigma}(z\cdot w)G^{d}_{\nu}(w\cdot y)\sqrt {g(w)}\, dw\sqrt {g(z)}\, dz.\cr&&   \label{diagram34}
\end{eqnarray}
Next, to perform the integrals  we use Eq. (\ref{pio}):
\begin{eqnarray}
 F^{(1)}_{\lambda\,\nu}(x,y)&=& 
   \frac{ f(\nu,\kappa,\sigma) 
 G^{d}_{\nu}(x\cdot y)}{(\nu^2-\lambda^2)}+   \frac{ f(\lambda,\kappa,\sigma)G^{d}_{\lambda}(x\cdot y)}{(\lambda^2-\nu^2)}  
 \cr &+&
 \sum_{k=0}^\infty
 \frac{ \rho(k;\kappa,\sigma)G^{d}_{\bu +2k  +\n+\sigma}(x\cdot y) }{((\bu +2k  +\n+\sigma)^2-\nu^2)(\bu +2k  +\n+\sigma)^2-\lambda^2)} ,
\end{eqnarray} 
where
\begin{align}
    f(\lambda,\kappa,\sigma) =\frac 1{2\lambda} (F_{tad}(\bu+\n+\sigma+\lambda)-F_{tad}(\bu+\n+\sigma-\lambda) ).
\end{align}

\section{Two-loop banana integrals: the watermelon}
\label{wmelon}
Here we provide a general formula for the watermelon with three lines and three independent mass parameters
\begin{align}
& I_3(\lambda,\nu,\kappa,d)=
  \frac {2\pi ^{\frac {d}{2}}}{\Gamma \left({\frac {d}{2}}\right)} \int _1^\infty G^{d}_{\lambda}(u )  G^{d}_{\nu}(u )  G^{d}_{\kappa}(u )(u^2-1)^{\frac{d-2}2}du.\label{bub3} 
  \end{align}
 The evaluation is made possible and actually rendered elementary by the  K\"all\'en-Lehmann formula (\ref{kl}); we use it to replace the first two factors in the integral with a series (\ref{bub3}): 
  \begin{align}
 I_3(\lambda,\nu,\kappa,d) &=  \sum_{k=0}^\infty\rho(k;\lambda,\nu)\, \frac {2\pi ^{\frac {d}{2}}}{\Gamma \left({\frac {d}{2}}\right)}  \int _1^\infty G^d_{\bu+2k  +\lambda+\nu}(u) G^{d}_{\kappa}(u )du \cr  &=  \sum_{k=0}^\infty\rho(k;\lambda,\nu)\,I_2\left(\left(\bu+2k  +\lambda+\nu\right),\kappa,d\right) \cr &= I_3^{(1)}(\lambda,\nu,\kappa,d)-I_3^{(2)} (\lambda,\nu,\kappa,d). \label{bub3bis}
  \end{align}
 where we used again the notation 
$
    \delta= \frac{d-1}{2}.
$ By inserting  the 1-loop integral  (\ref{bubads}) $I_3^{(1)}$ and $I_3^{(2)}$ are determined by summing the relevant hypergeometric series: 
 \begin{align}
&I_3^{(1)}(\lambda,\nu,\kappa,d) =  \frac{ \Gamma \left(1-\frac{d}{2}\right) }{2^{d} \pi ^{\frac{d}{2}}} \frac{\Gamma
   \left(\frac{d-1}{2}+\kappa \right)}{\Gamma
   \left(\frac{3-d}{2}+\kappa
   \right)}\sum_{k=0}^\infty\, \frac{ \rho(k;\lambda,\nu) }{((\frac{d-1}2+2k  +\lambda+\nu)^2 -\kappa^2 ) } \cr 
   &=\frac{2^{-2 \delta -3} \pi ^{-2 \delta -\frac{1}{2}} \Gamma \left(\frac{1}{2}-\delta
   \right) \Gamma (\delta +\kappa ) \Gamma (\delta +\lambda ) \Gamma (\delta +\nu )
   \Gamma (\delta +\lambda +\nu ) \Gamma (2 (\delta+\lambda +\nu )+1)}{\Gamma
   (\lambda +1) \Gamma (\nu +1) \Gamma (1-\delta +\kappa )  ((\delta  +\lambda +\nu )^2-\kappa^2) \Gamma (2 (\delta +\lambda +\nu )) \Gamma (2
   \delta +\lambda +\nu )} \times \cr &
\times \,     _9F_8\left(\begin{array}{l}\delta ,\delta +\lambda ,\delta +\nu ,\delta +\lambda +\nu ,\frac{1+\lambda +\nu }{2}
   ,\frac{2+\lambda +\nu }{2} ,\frac{2+\delta +\lambda +\nu
   }{2} ,\frac{\delta -\kappa +\lambda +\nu }{2} ,\frac{\delta +\kappa +\lambda
   +\nu}{2}  \cr 1+\lambda ,1+\nu ,1+\lambda +\nu ,\frac{\delta +\lambda +\nu }{2} ,\frac{2 \delta +\lambda
   +\nu }{2} ,\frac{1+2 \delta +\lambda +\nu }{2} ,\frac{2+\delta -\kappa +\lambda
   +\nu }{2} ,\frac{2+\delta +\kappa +\lambda +\nu }{2} \end{array};1\right),\cr & \label{i3}
\end{align}
and
\begin{align}
& I_3^{(2)}(\lambda,\nu,\kappa,d) =  \frac{ \Gamma \left(1-\frac{d}{2}\right) }{2^{d} \pi ^{\frac{d}{2}}} \sum_{k=0}^\infty\, \frac{ \rho(k;\lambda,\nu) }{((\frac{d-1}2+2k  +\lambda+\nu)^2 -\kappa^2 ) } 
\frac{\Gamma
   \left({d-1}+2k  +\lambda+\nu \right)}{\Gamma
   \left(1+2k  +\lambda+\nu).
   \right)} 
  \cr  &  = 
 \frac{4^{-\delta -1} \pi ^{-2 \delta -\frac{1}{2}} \Gamma \left(\frac{1}{2}-\delta
   \right) \Gamma (\delta +\lambda ) \Gamma (\delta +\nu ) \Gamma (\delta +\lambda +\nu
   +1)}{\Gamma (\lambda +1) \Gamma (\nu +1) \Gamma (\lambda +\nu +1) (\delta -\kappa
   +\lambda +\nu ) (\delta +\kappa +\lambda +\nu )} \times\cr &
 \times  \, _7F_6\left(\begin{array}{l}\delta ,\delta +\lambda ,\delta +\nu ,\delta +\lambda +\nu ,\frac{2+\delta +\lambda +\nu }{2} ,\frac{\delta -\kappa +\lambda +\nu }{2}
   ,\frac{\delta +\kappa +\lambda +\nu }{2} \cr 1+\lambda ,1+\nu ,1+\lambda +\nu
   ,\frac{\delta +\lambda +\nu }{2} ,\frac{2+\delta -\kappa +\lambda +\nu }{2} ,\frac{2+\delta +\kappa +\lambda
   +\nu }{2} \end{array};1\right),
   \end{align}

These expressions can be simplified by using a simple hypergeometric identity which, however, does not seem not to appear in the literature; we provide a derivation here.
By using the following identity for Pochhammer symbols
\begin{align}
    \frac {(a)_n (b)_n ((a+b)/2+1)_n}{(a+1)_n (b+1)_n ((a+b)/2)_n}=\frac {ab}{a+b}\left( \frac 1{a+n}+\frac 1{b+n} \right),
\end{align}
we can write
\begin{align*}
   {}_{p+1} F_p & \left(\begin{array}{l} a_1\ldots, a_{p-2}, \frac {a+b}2+1, a, b  \cr  b_1 ,\ldots,b_{p-3}, \frac {a+b}2, a+1, b+1 \end{array};1 \right)=\cr
    =&\sum_{n=0}^\infty \frac {(a_1)_n\cdots (a_{p-2})_n}{(b_1)_n\cdots (a_{p-3})_n} \frac 1{n!} \frac {(a)_n (b)_n ((a+b)/2+1)_n}{(a+1)_n (b+1)_n ((a+b)/2)_n}\cr
    =&\sum_{n=0}^\infty \frac {(a_1)_n\cdots (a_{p-2})_n}{(b_1)_n\cdots (a_{p-3})_n} \frac 1{n!} \frac {ab}{a+b}\left( \frac 1{a+n}+\frac 1{b+n} \right)\cr
    =&\sum_{n=0}^\infty \frac {(a_1)_n\cdots (a_{p-2})_n}{(b_1)_n\cdots (a_{p-3})_n} \frac 1{n!} \frac {1}{a+b}\left( b\frac {(a)_n}{(a+1)_n}+a\frac {(b)_n}{(b+1)_n} \right),
\end{align*}
which gives the identity
\begin{align}
   {}_{p+1}F_p \left(\begin{array}{l} a_1\ldots, a_{p-2}, \frac {a+b}2+1, a, b  \cr  b_1 ,\ldots,b_{p-3}, \frac {a+b}2, a+1, b+1 \end{array};1 \right)= \frac {b}{a+b}\ {}_{p-1} F_{p-2} \left(\begin{array}{l} a_1\ldots, a_{p-2}, a  \cr  b_1 ,\ldots,b_{p-3}, a+1 \end{array};1 \right)\cr
    +\frac {a}{a+b}\ {}_{p-1}F_{p-2} \left(\begin{array}{l} a_1\ldots, a_{p-2}, b  \cr  b_1 ,\ldots,b_{p-3}, b+1 \end{array};1 \right). \label{reduction}
\end{align}
By using \eqref{reduction} with $a=\frac {\delta+\lambda+\nu-\kappa}2$ and $b=\frac {\delta+\lambda+\nu+\kappa}2$ we get
\begin{align}
 &I_3^{(1)}(\lambda,\nu,\kappa,d) =   \frac{2^{-2 \delta -3} \pi ^{-2 \delta -\frac{1}{2}} \Gamma \left(\frac{1}{2}-\delta
   \right) \Gamma (\delta +\kappa ) \Gamma (\delta +\lambda ) \Gamma (\delta +\nu )
   \Gamma (\delta +\lambda +\nu )}{\Gamma
   (\lambda +1) \Gamma (\nu +1) \Gamma (1-\delta +\kappa )   \Gamma (2
   \delta +\lambda +\nu )} \times \cr 
  & \times \sum_{\epsilon=\pm}\frac 1{\delta  +\lambda +\nu+\epsilon\kappa}
   {}_7F_6\left(\begin{array}{l}\delta ,\delta +\lambda ,\delta +\nu ,\delta +\lambda +\nu, \frac {1+\lambda+\nu}{2} , \frac{2 +\lambda +\nu
   }{2} ,\frac{\delta +\epsilon\kappa +\lambda +\nu }{2}  \cr 
   1+\lambda ,1+\nu ,1+\lambda +\nu , \frac{2 \delta +\lambda
   +\nu }{2} ,\frac{1+2 \delta +\lambda +\nu }{2} ,\frac{2+\delta +\epsilon\kappa +\lambda
   +\nu }{2} \end{array};1\right),\cr \label{AA1}
\end{align}
and
\begin{align}
& I_3^{(2)}(\lambda,\nu,\kappa,d) = \frac{4^{-\delta -1} \pi ^{-2 \delta -\frac{1}{2}} \Gamma \left(\frac{1}{2}-\delta
   \right) \Gamma (\delta +\lambda ) \Gamma (\delta +\nu ) \Gamma (\delta +\lambda +\nu)}{\Gamma (\lambda +1) \Gamma (\nu +1) \Gamma (\lambda +\nu +1)} \times   \cr
  & \times  \, \sum_{\epsilon=\pm}\frac 1{\delta +\epsilon\kappa +\lambda +\nu }{}_5F_4\left(\begin{array}{l}\delta ,\delta +\lambda ,\delta +\nu ,\delta +\lambda +\nu ,\frac{\delta +\epsilon\kappa +\lambda +\nu }{2} \cr 1+\lambda ,1+\nu ,1+\lambda +\nu,\frac{2+\delta +\epsilon\kappa +\lambda
   +\nu }{2} \end{array};1\right). \label{AA2}
\end{align}
In the one-dimensional case (and, in general,  in odd negative dimension)  the above formulae become extremely simple: for example
 \begin{align}
& I_3^{(1)}(\lambda,\nu,\kappa,  1) = \frac{\lambda +\nu }{4 \kappa  \lambda  \nu  (\lambda +\nu-\kappa  ) (\lambda +\nu +\kappa )}, \cr
& I_3^{(2)} (\lambda,\nu,\kappa,1)=\frac{1}{4 \lambda  \nu  (\lambda +\nu-\kappa  ) (\lambda +\nu +\kappa)},
\cr &
I_3(\lambda,\nu,\kappa,  1)= \int_0^{\infty } \frac{e^{-t \lambda } e^{-t \nu }
    e^{-t \kappa }}{(2 \lambda ) (2 \nu
   ) (2 \kappa )} \, dt =I_3^{(1)}-I_3^{(2)} =\frac{1}{4 \kappa  \lambda  \nu  (\kappa +\lambda +\nu )}.
  \end{align}
  The watermelon diverges, as in flat space, starting at $d=3$. Actually, $I_3^{(2)}$ diverges at $d=3$ while the first divergence of $I_3^{(1)}$ is at $d=4$. At $d=2$ the difference  $I_3^{(1)}-I_3^{(2)}$ has a finite limit.

  \section{The effective potential for the $O(N)$ model}\label{sec5}
In this final section, as an application of our results, we compute the effective potential for the $O(N)$ scalar model on Anti de Sitter manifold in four dimensions. For generalities about the model we refer to the companion paper \cite{HSUdS}. As shown there, the 1-loop effective potential has the form $V_{\makebox{\em eff}}(\nu_0,\nu_1)=W(\nu_0)+W(\nu_1)$, where here
\begin{align}
    \nu_0=& \sqrt{m_R^2+3c_g \varphi_R^2+\frac 94}, \\
    \nu_1=& \sqrt{m_R^2+c_g \varphi_R^2+\frac 94},
\end{align}
and $W(\nu)$ satisfies
\begin{align}
    \frac d{d\nu^2} \frac d{d\nu^2} W(\nu)=-\frac 12 I_2(\nu,\nu,d),
\end{align}
by definition (notice that differently from the de Sitter case, here $I$ depends on $\nu^2$ through $\nu$, so the derivative in $\nu^2$ has to be intended as $\frac 1{2\nu} \frac d{d\nu}$).
Here we have already introduced the renormalized mass $m_R$ and field $\varphi_R$.
Integrating twice in $d\nu^2$ with $I$ given by \eqref{Ilambdanu4}, and $d=4-2\varepsilon$, we get
\begin{align}
    W(\nu)=&C\nu^2+D-\frac 1{64\pi^2} \left( \frac 1\varepsilon +1-\gamma +\log (4\pi) \right)\nu^4 +\frac {(\nu^2-1/4)^2}{32\pi^2} \psi\Big(\nu-\frac 12\Big)\cr
    &-\frac 1{32\pi^2} \int \Big( \nu^2-\frac 14\Big)^2 \psi'\Big(\nu-\frac 12\Big) d\nu,
\end{align}
where we have also included an integration by part, and $C$ and $D$ are integration constants. 
To evaluate the last integral, we use the Abel-Plana formula \cite{AbelPlana} to write 
\begin{align}
   \psi'\Big(\nu-\frac 12\Big)=\frac 12 \frac 1{\Big(\nu-\frac 12\Big)^2} +\frac 1{\nu-\frac 12} +4\int_0^\infty \frac {\nu-\frac 12}{\Big(\Big(\nu-\frac 12\Big)^2+t^2\Big)^2}\frac {tdt}{e^{2\pi t}-1},
\end{align}
which allows us to write
\begin{align}
   \int \Big( \nu^2-\frac 14\Big)^2 \psi'\Big(\nu-\frac 12\Big) d\nu=& \frac {\nu^4}4+\frac {\nu^3}3+\frac 5{24}\nu^2+\frac \nu4-\frac 7{48} +\frac {1-2\nu^2}{12} \log \Big(\nu-\frac 12\Big)^2\cr
   & -\mathcal B(\nu),
\end{align}
where 
\begin{align}
  \mathcal B(\nu):=& 2\int_0^\infty \frac {t^3 dt}{e^{2\pi t}-1} \frac {t^2-2\nu}{t^2 +\Big(\nu-\frac 12\Big)^2} 
  +12 \int_0^\infty \frac {t^2 dt}{e^{2\pi t}-1} \arctan \frac {\nu-\frac 12}t\cr
  &-2(1-2\nu^2) \int_0^\infty \frac {t dt}{e^{2\pi t}-1} \log \left( 1+\frac {t^2}{\Big(\nu-\frac 12\Big)^2} \right).
\end{align}
Notice that for $\nu\to\infty$ we have
\begin{align}
   \mathcal B(\nu)=\frac 3{2\pi^2} \zeta(3)+\frac 1{60}+\ldots, 
\end{align}
where the dots stay for terms that vanish for $\nu\to\infty$.\\
Adding the three level bare potential finally gives the renormalized potential
\begin{align}
    V_R=&\frac {m_R^2}2 \varphi_R^2 +\frac {c_g}4 \varphi_R^4 +\frac 1{32\pi^2} (m_R^2+3c_g \varphi_R^2+2)^2 \psi \left(\nu_0 -\frac 12\right)\cr
    &+\frac {N-1}{32\pi^2} (m_R^2+c_g \varphi_R^2+2)^2 \psi \left(\nu_1 -\frac 12\right) +\frac 1{32\pi^2} \mathcal B(\nu_0)+\frac {N-1}{32\pi^2} \mathcal B(\nu_1)\cr
    &-\frac1{32\pi^2}\Big(\frac {\nu_0^3}3+\frac 5{24}\nu_0^2+\frac {\nu_0}4-\frac 7{48} +\frac {1-2\nu_0^2}{12} \log \Big(\nu_0-\frac 12\Big)^2\Big)\cr
    &-\frac{N-1}{32\pi^2}\Big(\frac {\nu_1^3}3+\frac 5{24}\nu_1^2+\frac {\nu_1}4-\frac 7{48} +\frac {1-2\nu_1^2}{12} \log \Big(\nu_1-\frac 12\Big)^2\Big)\cr
    & -\frac 1{64\pi^2} \left[(m_R^2+3c_g \varphi_R^2+2)^2+ (N-1)(m_R^2+c_g \varphi_R^2+2)^2 \right] \log {\mu_R^2}. \label{potenziale adimensionale}
\end{align}
We can now restore full dimensions by reintroducing the curvature radius $R$, such that $\Lambda=-\frac 3{R^2}$ is the cosmological constant and 
\begin{align}
    V=\frac V{R^4}, \qquad m_R=Rm, \qquad \varphi_R=R\varphi, \qquad \mu_R=R\mu.
\end{align}
This gives for the effective potential in $d=4$:
\begin{align}
    V=&\frac {m^2}2 \varphi^2 +\frac {c_g}4 \varphi^4 +\frac 1{32\pi^2} \Big(m^2+3c_g \varphi^2-\frac 23 \Lambda\Big)^2 \psi \left(\nu_0 -\frac 12\right)\cr
    &+\frac {N-1}{32\pi^2} \Big(m^2+c_g \varphi^2-\frac 23\Lambda\Big)^2 \psi \left(\nu_1 -\frac 12\right) +\frac {\Lambda^2}{288\pi^2} \mathcal B(\nu_0)+\frac {(N-1)\Lambda^2}{288\pi^2} \mathcal B(\nu_1)\cr
    &-\frac {\Lambda^2}{288\pi^2}\Big(\frac {\nu_0^3}3+\frac 5{24}\nu_0^2+\frac {\nu_0}4-\frac 7{48} +\frac {1-2\nu_0^2}{12} \log \Big(\nu_0-\frac 12\Big)^2\Big)\cr
    &-\frac{(N-1)\Lambda^2}{288\pi^2}\Big(\frac {\nu_1^3}3+\frac 5{24}\nu_1^2+\frac {\nu_1}4-\frac 7{48} +\frac {1-2\nu_1^2}{12} \log \Big(\nu_1-\frac 12\Big)^2\Big)\cr
    & -\frac 1{64\pi^2} \left[\Big(m^2+3c_g \varphi^2-\frac 23\Lambda\Big)^2+ (N-1)\Big(m^2+c_g \varphi^2-\frac 23\Lambda\Big)^2 \right] \log {\frac {3\mu^2}{|\Lambda|}}, \label{potenziale}
\end{align}
with
\begin{align}
    \nu_0=& 3\sqrt{\frac 14 -\frac {m^2}{3\Lambda}-c_g\frac {\varphi}{\Lambda}}, \\
    \nu_1=& 3\sqrt{\frac 14 -\frac {m^2}{3\Lambda}-c_g\frac {\varphi}{3\Lambda}}.
\end{align}
For small cosmological constant we get the expansion
\begin{align}
    V=&\frac {m^2}2 \varphi^2 +\frac {c_g}4 \varphi^4 +\frac 1{64\pi^2} \Big(m^2+3c_g \varphi^2\Big)^2 \log \frac {m^2+3c_g \varphi^2}{\mu^2}\cr
    &+\frac {N-1}{64\pi^2} \Big(m^2+c_g \varphi^2\Big)^2 \log \frac {m^2+c_g \varphi^2}{\mu^2}  \cr
    &-\frac 1{8\pi^2} \left( \frac {m^2+3c_g \varphi^2}3\right)^{\frac 32} \sqrt{-\Lambda}-\frac {N-1}{8\pi^2} \left( \frac {m^2+c_g \varphi^2}3\right)^{\frac 32} \sqrt{-\Lambda}\cr
    &-\frac {\Lambda (m^2+3c_g \varphi^2)}{288\pi^2} \left(\frac {11}8 +\frac 12 \log \frac {3m^2+9c_g \varphi^2}{-\Lambda} +6 \log \frac {m^2+3c_g \varphi^2}{\mu^2} \right)\cr
    & -\frac {\Lambda (N-1) (m^2+c_g \varphi^2)}{288\pi^2} \left(\frac {11}8 +\frac 12 \log \frac {3m^2+3c_g \varphi^2}{-\Lambda} +6 \log \frac {m^2+c_g \varphi^2}{\mu^2} \right)\cr
    & +O(|\Lambda|^{\frac 32}). \label{potenziale}
\end{align}
Notice that there is an important difference as compared to the effective potential in de Sitter computed in \cite{HSUdS}: here the cosmological function appears as the square root $\sqrt{-\Lambda}$ in place of $\Lambda$. This is due to the symmetry   $\nu\to-\nu$ of the Wightman function in  the de Sitter case (that is a symmetry of Legendre functions of the first kind) a symmetry that AdS quantum fields do not share.

\section{Conclusions}
In the present paper, we have revisited methods and tools for quantum field theory on the anti de Sitter spacetime already available in the literature \cite{bem,{begmp2}} but whose potential seems to have gone unnoticed. In particular, we have shown here their power in computing loop integrals directly, showing no need of bootstrapping at this level. 

We have also shown the technical superiority of conceiving the correlation functions of anti de Sitter quantum fields as Legendre functions of the second kind rather than the less specific, altough equivalent, hypergeometric functions. 

The  methods used in this paper include theorems that render almost elementary the computation of certain integrals which otherwise quite complicated even in flat Minkowski spacetime: one example is the two-loop banana integral, the watermelon, with three arbitrary masses.  The study of the watermelon where it diverges at $d= 3, 4, 5$ dimensions has been left for future study where we will also show how to deal with loop Witten diagrams, i.e. loop integrals with external legs going to spacelike infinity.

\acknowledgments{U.M. gratefully  thanks the Department of Theoretical Physics of CERN for hospitality and support while writing  this paper.}


\newpage
\appendix

\begin{appendix} \label{aapa}
\section{Remarks on the various spectral representations of dS and AdS propagators}
In this section we would like to discuss some points on the ongoing literature on harmonic analysis on dS and AdS, which, in our opinion, need clarification.

In section {\ref{anal} we have shown  that {\bf\em Legendre's associate functions of the second kind}\footnote{More precisely, Legendre's functions of the second kind multiplied by the factor $(\zeta^2-1)^{-\frac{d-2}4}$. \label{fn} }   have analyticity properties that are tailor-made for the geometry of the anti de Sitter manifold and its covering: the maximal analyticity property  of the two-point functions (\ref{s.9}) in  the covering cut-plane $\wt \Delta_1$
is a consequence  of  AdS invariance, local commutativity and  the positivity of the spectrum of the energy operator $M_{0d}$;   points of the cut of $\wt \Delta_1$ correspond to pair of timelike separated events on the real manifold $\wt{AdS}_d$.

On the other hand {\bf\em Legendre's functions of the first kind}\footnote{ See footnote \ref{fn}.}   have analyticity properties that are tailor-made for the geometry of the de Sitter manifold.
A short summary will help in making clear what follows. The complex de Sitter universe may be represented as  the  one-sheeted hyperboloid 
\beq
dS_d^{(c)} = \{z \in M_{d+1}^{(c)}\ :\ z\cdot z = -R^2=-1\}\ .
\label{s.2}\endq
immersed in a $(d+1)$-dimensional complex Minkowski spacetime.
The same letter $\zeta$ used  before  here 
denotes the complex de Sitter invariant   variable
\begin{equation}\zeta = z_1\cdot z_2  = z^0_1 z^0_2-z^1_1 z^1_2 - \ldots -z^d_1 z^d_2,
\end{equation}
 i.e. the scalar  product in the ambient spacetime sense of  two complex events $z_1, z_2 \in dS_d^{(c)}$.
 Two real events $x_1$ and $x_2$ are timelike separated if and only if 
\begin{equation}
(x_1-x_2)^2  = - 2 - 2 x_1\cdot x_2 >0. \label{causal}
\end{equation}
The canonically normalized (Bunch-Davis) maximally analytic two-point function of a de Sitter Klein-Gordon field 
is given by
\begin{align}
& W^{(dS)}_\kappa (z_1,z_2)   = w^{(dS)}_\kappa(\zeta)=  {\Gamma\left(\frac{d-1}2 + i \kappa\right)\Gamma\left(\frac{d-1}2 - i \kappa\right)   \over 2 (2\pi)^{d/2}}
(\zeta^2-1)^{-{d-2 \over 4}}\,P_{ -{1\over 2}+ i \kappa }^{-{d-2 \over 2}}(\zeta)= 
 \label{pppds}\\ &= 
{\Gamma\left(\frac{d-1}2 + i \kappa\right)\Gamma\left(\frac{d-1}2 - i \kappa\right) \over
(4\pi)^{d/2}\Gamma\left({d\over 2}\right)}
{}_2F_1\left(\frac{d-1}2 + i \kappa,\frac{d-1}2 - i \kappa;\ {d\over 2};\ {1-\zeta\over 2}\right )
\label{wigds}
\end{align}
where the (complex) mass squared is related to the parameter $\kappa$ and the spacetime dimension $d$ as follows (we set $R=1)$:
\begin{equation}
m^2  = \frac {(d-1)^2}{4} +\kappa^2. \label{cmass}
\end{equation}

The Legendre function \cite{bateman} of the first kind  $P^{ \alpha }_{\beta}(\zeta) $  is  holomorphic in the cut-plane  
$\Delta_2$ -- the same as $Q^{ \alpha }_{\beta}(\zeta) $ --  but the reduced two-point function  
$w^{(dS)} _\kappa(\zeta)$ 
is holomorphic in the  larger domain 
\beq \zeta  = z_1\cdot z_2 \in \Delta = \bC\setminus (-\infty,\ -1],     
\label{r.1.2}\endq 
i.e. everywhere except on the causality cut (\ref{causal})  cut ({\em maximal analyticity property},  \cite{bm}). The factor $(\zeta^2-1)^{-{d-2 \over 4}}$ exactly compensates the singularity of  $P_{ -{1\over 2}+ i \nu }^{-{d-2 \over 2}}(\zeta) $ at $\zeta=1$ making the reduced two-point function regular there.
\vskip 10 pt

All the above being said and having specified the differences  between the  de Sitter and the anti-de Sitter two-point functions, let us focus now on the mathematical  possibility to express the anti-de Sitter {\em reduced} two-point function (\ref{s.9}) as a superposition of {\em reduced} de Sitter two-point functions (\ref{pppds}) i.e. as a  superposition of Legendre functions of the first kind.

The first thing we would like to recall here is that a Fourier-like representation of the  two-point function (\ref{pppds}) (equivalently (\ref{wigds})) in terms of {\em plane waves}  is available since 1994   \cite{bgm,bm}:
\begin{align}
W^{(dS)}_\nu (z_1,z_2)   &= \frac{\Gamma\left(\frac{d-1}2 + i \kappa\right)\Gamma\left(\frac{d-1}2 - i \kappa\right)
e^{\pi\kappa}}{2^{d+1}\pi^d}
\int_{\gamma}(\xi\cdot z_1)^{-\frac{d-1}2 - i \kappa}(\xi\cdot z_2)^{-\frac{d-1}2 + i \kappa}\,\alpha(\xi).  \label{uup}
\end{align}
Here $\xi\in C^+=\partial V^+= \{\xi\in M_{d+1}, \ \xi^2=0,\ \xi^0>0\}$  is a forward light-like vector in the ambient spacetime; it can be also thought as a point of the  boundary  of the de Sitter manifold at timelike infinity. A natural physical interpretation  for such a $\xi$ is that of the asymptotic direction of the momentum of a particle \cite{cacciatori2008}.

The plane waves $ \psi_\lambda(z,\xi) = \left({ \xi \cdot z}\right)^{\lambda} $ composing the above formula solve individually the Klein-Gordon equation in the respective tuboids:
\begin{equation} \Box_{dS} \psi (z)   +m^2 \psi (z) =0,\label{KGnu}\end{equation}
where 
\begin{equation}
 m^2 = -\lambda(\lambda+d-1) = \frac {(d-1)^2}{4} +\kappa^2, \ \ \  \lambda=-\frac{d-1}2 + i \kappa,
\end{equation}
($d$ is the spacetime dimension).

The complex de Sitter events $z_1$ and $z_2$ belong  to the backward and, respectively, the forward de Sitter tuboids  $\TT_-$ ands  $\TT_+$; these are the intersections of the ambient tubes   $T_-$ ands  $T_+$ (which are related to the positivity of the energy momentum operator \cite{Streater}) with the complex de Sitter manifold:
\beq
\TT_\pm = \{x+iy \in X_d^{(c)}\ :\  x\cdot y = 0 ,\ \ y^2>0, \ \text{sign}\  y^0 = \pm \}\ .
\label{s.2.1}\endq
In  standard coordinates, the ($d-1$)-form $\alpha(\xi)$  in Eq. (\ref{uup}) is written 
\begin{equation} \alpha(\xi) = (\xi^0)^{-1} \sum_{j=1}^d (-1)^{j+1}
\xi^j\,d\xi^1\ldots\ \wh{d\xi^j}\ldots\ d\xi^d ;
\label{f.16}\end{equation}
$\gamma$
denotes any
($d-1$)-cycle in the forward
light-cone $C^+$; the integrand in Eq. (\ref{uup}) is a closed differential form and therefore  the result does not depend on the choice of  the manifold  $\gamma$.  In particular we may choose the  unit spherical section of the cone 
$\sd$ (equipped with its canonical orientation):
\begin{equation}
 \gamma_0=\sd= C^+ \cap \{\xi\ :\ \xi^0 =1\}
=  \{ \xi \in C^+ : {\xi^1}^2 + \ldots + {\xi^d}^2 =
1\}.
\end{equation}
With this choice $\alpha(\xi)$  coincides
with the rotation invariant measure $d\xi$ on  $\sd$ normalized as
usual:
\begin{equation}
\omega_{d}=\int_{\gamma_0}\alpha(\xi)  = \frac{2\pi^\frac d2}{\Gamma\left(\frac d2\right)}.
\label{norms}
\end{equation}
Eq. (\ref{uup}) is very flexible and may be used to produce many distinct integral representation of the Legendre function of the first kind. 
One possible construction goes as follows: the tubes  $\TT_-$ and   $\TT_+$ are bordered by two Riemannian (as opposed to Euclidean) manifolds  ${\cal H}_-$ and   ${\cal H}_+$ whose points are purely imaginary:
\beq
{\cal H}_\pm = \{x+iy \in X_d^{(c)}\ : x=0, \ \ y^2=R^2=1, \ \text{sign}\  y^0 = \pm \}\ .
\label{s.2.1}\endq
Both ${\cal H}_-$ and ${\cal H}_+$ are equivalent to the $d$-dimensional Lobachevsky space  $H_d$ which is also equivalent to the Euclidean AdS manifold, see Eq. $\Lobad$ (\ref{s.1}).

We now choose the two points in Eqs. (\ref{uup}) to be purely imaginary: $z_1=-iY_1\in {\cal H}_-$ and   $z_2=iY_2\in {\cal H}_+$; we  choose to integrate over  the parabolic basis of the cone
\begin{equation}
 \gamma_p= C^+ \cap \{\xi\ :\ \xi^0+\xi^d =1\},
\end{equation}
which we parametrize as follows:
\begin{equation}
\xi^0 = P^0 = \frac 12 +x^2 ,  \ \ \  \xi^i = P^i = x^i, \ \ \  \xi^d =P^d =\frac 12 - x^2, \ \ \  dP = dx
\end{equation}
and the result is\footnote{In a portion of the recent literature concerning harmonic analysis on the anti-de Sitter manifold a formula like (\ref{uupom})  is called 
a "split representation"   (see e.g. \cite{pene2} and reference therein).
The name has been introduced to denote  "bulk-to-bulk propagators" that are represented as 
integrals of products of two scalar "bulk-to-boundary propagators". 
Our simple construction shows that 
"bulk-to-boundary propagators" are nothing but plane waves and a split representation is a special instance 
of the general analytic formula (\ref{uup}). This remains true also in the more complicated tensorial cases treated in \cite{pene2}.
We believe that our old denomination  \cite{bm} "Fourier-like" or "plane-waves representation"  makes more justice to Eq. (\ref{uup}) and to any  other split representation. }
\begin{align}
\Omega_\kappa (Y_1,Y_2) = W^{(dS)}_\kappa (-iY_1,iY_2)   &= N(\kappa)
\int_{\gamma}(P \cdot Y_1)^{-\frac{d-1}2 - i \kappa}( P \cdot Y_2 )^{-\frac{d-1}2 + i \kappa}\,dP  \label{uupom}
\end{align}
(see \cite[Eq. 113]{Pene}). 

Even though it is a stretch, we might now insist in identifying  both ${\cal H}_-$ and ${\cal H}_+$ with the same copy of a Euclidean AdS manifold;  in the same vein we may also, identify the light-cone $C^+$ with  the boundary of the Euclidean AdS manifold; with the these premises  $\Omega_\kappa (Y_1,Y_2)$ may be taken as a basis
of harmonic functions (proportional to Legendre functions of the first kind, see Eq. (\ref{pppds})) that can be used to develop the AdS propagators. While this may be a useful technical tool, as for instance in the calculations of Witten diagrams in \cite{Pene}, it is somehow an unnatural move as we try to explain now.

Expressing the dS two-point function in as a superposition of  AdS correlators we already did in  Eq.  (\ref{pop}) which we we now rewrite in a more suggestive way:
\begin{equation}
w^{AdS}_{-\nu} (\zeta)  - w^{AdS}_{\nu} (\zeta)  = \sin \pi\nu \ w^{dS}_{i\nu} (\zeta) . \label{pop2}
\end{equation}

Expressing the scalar AdS propagator  (which is  a  Legendre functions of the second kind) as a  superposition of Legendre functions of the first kind is also simple; it
can be done in a one-line calculation by taking a generalized  Mehler-Fock inverse  transform \cite[Eq. 3.15.8]{bateman}. In the two-dimensional case $d=2$  
the formula we are looking for is in books since a very long time  \cite [Eq.  (7.213))]{grad}: 
\begin{eqnarray}
Q_{-\frac{1}{2} + \nu}(u)} =  \int_0^\infty \kappa \tanh (\pi \kappa) 
 P^{}_{-\frac{1}{2} + i \kappa} (u)\,
{(x^2+\nu^2)^{-1} \ dx , \ \ \ \Re \kappa >0, \ \ u \in \Delta_2\; \label{grad};
\end{eqnarray}
the above identity holds in particular for real $u>1$. It is however clear that something goes wrong for $-1<u<1$:  the  integrand is holomorphic for  $u\in \Delta$ while the Legendre function at the l.h.s. is defined only  for $u\in \Delta_2\subset \Delta$; the cut of $Q$ is longer and the integral ill-defined on the cut.

To find the extension of the above formula to general dimension $d$, let us  take the Riemann-Liouville transform of both sides of Eq. (\ref{grad}):
\begin{align}
& \int_z^\infty {  (u-z)^{\mu-1 }
 P_{-\frac{1}{2}+i \kappa
   }(u)} du = \cr & =\frac1 {\pi } 
   \cosh (\pi  x )  \Gamma (\mu )
   \Gamma \left(\frac{1}{2}-\mu -i
   \kappa \right) \Gamma
   \left(\frac{1}{2}-\mu +i \kappa
   \right) \left(z^2-1\right)^{\mu /2} P_{-\frac{1}{2}+i \kappa
   }^{\mu }(z), \cr 
& \int_z^\infty {  (u-z)^{\mu-1 }
 Q_{-\frac{1}{2}+\nu
   }(u)} du =  \cr & = \frac 1 {\pi }
   e^{-i \pi  \mu } \cos (\pi 
   \mu + \pi  \nu ) \Gamma (\mu )
   \Gamma \left(\frac{1}{2}-\mu -\nu \right) \Gamma
   \left(\frac{1}{2}-\mu +\nu
   \right) \left(z^2-1\right)^{\mu
   /2} Q_{-\frac{1}{2}+\nu }^{\mu
   }(z). \cr & 
\end{align}
The sought extension immediately follows:
\begin{align}
   Q_{-\frac{1}{2}+\nu }^{\mu
   }(z) & = 
   e^{i \pi  \mu } \int_0^\infty\frac{ \kappa\sinh (\pi \kappa) }{ (\kappa^2+\nu^2)}\frac{ \Gamma \left(\frac{1}{2}-\mu -i
   \kappa \right) \Gamma
   \left(\frac{1}{2}-\mu +i \kappa
   \right)}{\cos (\pi 
   \mu + \pi  \nu )
   \Gamma \left(\frac{1}{2}-\mu -\nu \right) \Gamma
   \left(\frac{1}{2}-\mu +\nu
   \right)} P^\mu_{-\frac{1}{2} + i \kappa} (z)d\kappa . 
\end{align}
Taking into account the identity  (\ref{wq.40}) 
we get our final formula generalizing (\ref{grad}):
\begin{align}
  e^{-i \pi  \frac{d-2}2 }  Q_{-\frac{1}{2}+\nu }^{\frac{d-2}2
   }(z) & = 
   \int_0^\infty\frac{ \kappa\sinh (\pi \kappa) }{\pi (\kappa^2+\nu^2)}{ \Gamma \left(\frac{d-1}{2} -i
   \kappa \right) \Gamma
   \left(\frac{d-1}{2} +i \kappa
   \right)}{} P^{-\frac{d-2}2}_{-\frac{1}{2} + i \kappa} (z)d\kappa. \label{a11}
\end{align}
Eqs. (\ref{s.9}) and (\ref{pppds}) allow to rewrite the above identity  (\ref{a11}) in a suggestive way:
\begin{align}
&w^{(AdS)}_{\nu}(\zeta) =
\frac 2 \pi \int_0^\infty\frac{  x\sinh (\pi \kappa) }{(\kappa^2+\nu^2)}w^{(dS)}_\kappa(\zeta)d\kappa \label{weird0}, \ \ \ \zeta\in 
\Delta_2. 
\end{align}
One should never forget that   $w^{(AdS)}_{\nu}(\zeta)$ and $w^{(dS)}_{\kappa}(\zeta)$ have different domains of analyticity. Even the variable $\zeta$ has different meanings at the rhs and at the lhs. In particular the locus of AdS coincident Euclidean points projects on $\zeta=1$ while the coincident dS Euclidean points project on  $\zeta=-1$. In these respective places the Schwinger functions have distributional singularities that in integer dimensions cause the appearance of a $\delta$ source at the r.h.s. of the Klein-Gordon equation. 

Inserting Eq. (\ref{uupom}) into  Eq. (\ref{weird0}) one gets the so-called "split representation" \cite{Costa} of the AdS Schwinger function:
\begin{align}
&w^{(AdS)}_{\n}(Y_1\cdot Y_2) =
\frac 2 \pi \int_0^\infty\frac{  \kappa\sinh (\pi \kappa) }{(\kappa^2+\nu^2)}N(\kappa)
\int_{\gamma}(P \cdot Y_1)^{-\frac{d-1}2 - i \kappa}( P \cdot Y_2 )^{-\frac{d-1}2 + i \kappa}\,dP \label{split}. 
\end{align}
This representation has proven to be useful in computing Witten diagrams on the Euclidean anti de Sitter space \cite{Pene}. However, {\em  it cannot be extended to the real anti de Sitter manifold}.

Finally we may use Eq. (\ref{pop})  and write 
\begin{align}
&w^{(AdS)}_{\nu}(\zeta) =
\frac i \pi \int_0^\infty\frac{  \kappa }{ (\kappa^2+\nu^2)}(w^{(AdS)}_{i\kappa} (\zeta)- w^{(AdS)}_{-i\kappa}(\zeta) )d\kappa \label{weird} \end{align}
But the important  formula is Eq. (\ref{a11}). The equivalent expressions (\ref{weird0}) and (\ref{weird}) are just ornamental.

}

The idea of expanding the AdS time-ordered propagator in Legendre functions of the first kind has been introduced long ago by D\"usedau and Freedman \cite{dusedau} as an intermediate step in computing the K\"all\'en-Lehmann expansion of a propagator. Even in the title of their paper, D\"usedau and Freedman  reserve the name of {\em spectral representation}
to the K\"all\'en-Lehmann expansion, the spectral parameter being the mass of the  two-point functions entering in there. 
Their Eq. (31)   corresponds to our Eq. (\ref{a11}) specialized to $d=4$ and is  
described  as a special instance of an integral transform of Gelfand and Graev \cite{dusedau}.

The possibility of representing AdS propagators  in terms of Legendre functions of the first kind  has been recently rediscovered  and exploited. In this recent literature this  is no more intended as an  intermediate step towards a true K\"all\'en-Lehmann representation of the propagator but as a "spectral representation" {\em per se}.

Eq. (\ref{weird0}) is  called in \cite[Appendix B.1]{Pene} "the harmonic space representation of the bulk to bulk propagator". Eq. (\ref{weird})  in   \cite[Appendix B.2]{Carmi2018} is  the "spectral representation" of the propagator. These denominations are, in our opinion, rather misleading because the Legendre functions of the first and second kind (obviously)  solve the same Legendre differential equation (the delta terms arising only at coincident Euclidean points) and also because that representation has already a name: generalized inverse Mehler-Fock  transformation (\ref{a11}).

The "spectral representation" is sometimes rather  inconvenient: a key result of \cite{Carmi2018} consists in determining (through a quite involved bootstrappism and a certain  amount of clever guesswork) the "spectral representation"  of the bubble $w_{\lambda}(\zeta)w_{\nu}(\zeta)$ 
in the particular case of equal mass parameters $\lambda= \nu$.

It is indeed quite elementary  to   compute such "spectral representation"  knowing the K\"all\'en-Lehmann representation (\ref{x.15}) and the Mehler-Fock antitransform of the propagator (\ref{weird0});
here we perform the calculation in the general case of two different masses $\lambda$ and $\nu$ and for the maximally analytic functions which include as a special case the Schwinger function at non-coinciding points:
\begin{align}
&w^{(AdS)}_{\lambda}(\zeta)\,w^{(AdS)}_{\nu}(\zeta) =
\frac 2 \pi \int_0^\infty \sum_{k=0}^\infty\frac{   \rho(k;\lambda,\nu) }{(\kappa^2+(\frac{d-1}2 +2k +\lambda+\nu)^2)} \kappa\sinh (\pi \kappa)w^{(dS)}_\kappa(\zeta)d\kappa .\label{kll} 
\end{align}
By taking the series first,  the weight is written as a generalized hypergeometric series:
\begin{align}
& B_d(\lambda,\nu,\kappa) = \sum_{k=0}^\infty\frac{    \rho(k;\lambda,\nu) }{(\kappa^2+(2k +\frac{d-1}2 +\lambda+\nu)^2)}.
\end{align}
We have already encountered this series when computing the watermelon in Sect. \ref{wmelon} (see Eq. (\ref{i3})): 
\begin{align}
& B_d(\lambda,\nu,x) =  \frac{\pi ^{-\delta } \Gamma (\delta +\lambda ) \Gamma (\delta +\nu ) \Gamma (\delta +\lambda +\nu +1)}{2 \Gamma (\lambda +1) \Gamma (\nu +1) \Gamma (2 \delta
   +\lambda +\nu ) \left((\delta +\lambda +\nu )^2+\kappa^2\right)} \cr &
\times \,     _9F_8\left(\begin{array}{l}\delta ,\delta +\lambda ,\delta +\nu ,\delta +\lambda +\nu ,\frac{1+\lambda +\nu }{2}
   ,\frac{2+\lambda +\nu }{2} ,\frac{2+\delta +\lambda +\nu
   }{2} ,\frac{\delta -i \kappa +\lambda +\nu }{2} ,\frac{\delta +i \kappa +\lambda
   +\nu}{2}  \cr 1+\lambda ,1+\nu ,1+\lambda +\nu ,\frac{\delta +\lambda +\nu }{2} ,\frac{2 \delta +\lambda
   +\nu }{2} ,\frac{1+2 \delta +\lambda +\nu }{2} ,\frac{2+\delta -i \kappa +\lambda
   +\nu }{2} ,\frac{2+\delta +i \kappa+\lambda +\nu }{2} \end{array};1\right).\cr & \label{ii3}
\end{align}
In the simplest case where the two masses are equal the above formula reduces to
\begin{align}
& B_d(\nu,\nu,\kappa) =  \frac{\pi ^{-\delta } \Gamma (\delta +\nu )^2 \Gamma (\delta +2 \nu +1)}{2 \Gamma (\nu +1)^2 \Gamma (2 (\delta +\nu )) \left((\delta +2 \nu )^2+\kappa^2\right)} \cr &
\times \,    \, _7F_6\left(\begin{array}{l}\delta ,\nu +\frac{1}{2},\frac{\delta }{2}+\nu +1,\frac{\delta }{2}+\nu -\frac{i\kappa}{2},\frac{\delta }{2}+\nu +\frac{i \kappa}{2},\delta
   +\nu ,\delta +2 \nu \cr \nu +1,\frac{\delta }{2}+\nu ,\frac{\delta }{2}+\nu -\frac{i \kappa}{2}+1,\frac{\delta }{2}+\nu +\frac{i \kappa}{2}+1,\delta +\nu +\frac{1}{2},2
   \nu +1\end{array};1\right)\label{ii3b}
\end{align}
By using formula \eqref{reduction} to simplify Eq.  (\ref{ii3b}) (with $a=\frac{\delta +2 \nu -i \kappa}2$ and $b=\frac{\delta +2 \nu +i \kappa}2$) we see that Eq. (4.7) of \cite{Carmi2018} is an immediate consequence of the K\"all\'en-Lehmann formula.

The weight  $B_d(\lambda,\nu,\kappa)$ of the "spectral representation" is proportional to $I_3^{(1)}(\lambda,\nu,i \kappa,d)$  which is one of the two terms of the 2-loop vacuum diagram studied in Sect. \ref{wmelon}: this  explains why $B_d(\lambda,\nu,x)$ may diverge at $d=3,4\ldots$; 
actually, $B_d(\lambda,\nu,\kappa)$  does not diverge at $d=3$ because the divergences in $I_3^{(1)}(\lambda,\nu,i \kappa,d)$ are milder than the divergences in $I_3^{(2)}(\lambda,\nu,i \kappa,d)$; the latter  does diverge also at $d=3$.

Whatever is the case, $B_d(\lambda,\nu,\kappa)$ is to be compared with the K\"all\'en-Lehmann weight (\ref{kl}), which is splendidly regular at $d=3,4\ldots$

In our opinion, this is already a sufficient reason to avoid the use of 
the so-called "spectral representation" and keep using the good old K\"all\'en-Lehmann representation.

Note also that in the "spectral representation"  
the bubble is expanded as  an integral of Legendre functions of the first kind as opposed to the series of Legendre functions of the second kind exhibited  at the r.h.s. of Eq. (\ref{x.15}).  The quasi-periodical character of the bubble is now completely hidden in a weird  representation. 
\vskip 10 pt
From a mathematical viewpoint,  the Mehler-Fock representation (\ref{a11}) of the Legendre functions may however be useful  in other situations, as for the calculation of Euclidean Witten diagrams in  \cite{Pene}. 
For example, using  the known   integral of three Legendre functions  of the first kind \cite{us}
\begin{eqnarray}
h_d(\lambda,\nu, \n) = \int_1^\infty
{P^{-\frac{d-2}{2}}_{-\frac{1}{2} +
i\lambda}(u)P^{-\frac{d-2}{2}}_{-\frac{1}{2} +
i\nu}(u)} P^{-\frac{d-2}{2}}_{-\frac{1}{2} + i \n} (u)\,
{(u^2-1)^{-\frac{d-2}4}} \ du =  \label{integralret}
\cr
=
\frac{2^{\frac d 2}}{(4\pi)^{\frac32}
\Gamma\left(\frac{d-1}{2}\right)} \frac{ \ \prod_{\epsilon,\epsilon',\epsilon''=\pm 1}
\Gamma\left(\frac{d-1}{4}
+\frac{i\epsilon\lambda+i\epsilon'\nu +i\epsilon'' \kappa}{2}\right)}
{ \prod_{\epsilon,\epsilon' \epsilon''=\pm 1}
\Gamma\left(\frac{d-1}{2}+i\epsilon \lambda\right)
\Gamma\left(\frac{d-1}{2}
+i\epsilon' \nu\right) \Gamma\left(\frac{d-1}{2}+i\epsilon'' \n\right)}
\label{theformula}
 \end{eqnarray}
we may compute the following related integral that may play a role in either dS or AdS two-loop calculations (details are omitted):
\begin{align}&
\int_1^\infty
{Q^{-\frac{d-2}{2}}_{-\frac{1}{2} +
a}(u)P^{-\frac{d-2}{2}}_{-\frac{1}{2} +
i\nu}(u)} P^{-\frac{d-2}{2}}_{-\frac{1}{2} + i \n} (u)\,
{(u^2-1)^{-\frac{d-2}4}} \ du 
\cr 
 &= \frac{ \pi ^{3/2} {
   G_{8,8}^{5,5}\left(1\left|
\begin{array}{c}
 1-\frac{a}{2},
 \frac{5-d}{4} + \frac{ i \kappa- i \nu}{2},
  \frac{5-d}{4} - \frac{ i \kappa- i \nu}{2},
 \frac{5-d}{4} + \frac{ i \kappa +i \nu}{2},
  \frac{5-d}{4} - \frac{ i \kappa+ i \nu}{2}
,0,\frac{1}{2},1+\frac{a}{2} \\
 \frac{a}{2},
\frac{d-1}{4} + \frac{ i \kappa}{2} -\frac{ i \nu}{2},
  \frac{d-1}{4} - \frac{ i \kappa}{2} +\frac{ i \nu}{2},
 \frac{d-1}{4} + \frac{ i \kappa}{2} +\frac{ i \nu}{2},
  \frac{d-1}{4} - \frac{ i \kappa}{2} -\frac{ i \nu}{2},\frac{1}{2},1,-\frac{a}{2} \\
\end{array}
\right.\right)}}{2^{2-\frac{d}{2}} e^{\frac{1}{2} i d
   \pi }\cos \left(\pi  a-\frac{\pi 
   d}{2}\right)\Gamma \left(\frac{d-1}{2}\right)\prod_{\epsilon=\pm} \Gamma
   \left(\frac{d-1}{2}+i \epsilon 
   \kappa \right) \Gamma
   \left(\frac{d-1}{2}+i \epsilon 
   \nu \right)\Gamma \left(\frac{d-1}{2}+ \epsilon 
   a \right)} \cr & 
\end{align}
where $G$ is Meijer's function. The limit $a\to i \lambda$ may be expressed as a combination  of hypergeometric series $_7F_6$ which we do not reproduce here.

\end{appendix}

\section{Appendix. Proof of theorem \ref{k-l4}}
\label{apkl4}
Let $(\bu,\ \lambda,\ \nu)\in V$, $\z\in \wh\Delta_1$, $k\in \bZ_+$.
Using (\ref{s.9}) and \cite[3.3 (44) pp. 136-137]{bateman}, we get
\beq
S_k(\bu,\ \lambda,\ \nu,\ \z) = A_k(\bu,\ \lambda,\ \nu)
B_k(\bu,\ \lambda,\ \nu,\ \z),
\label{zm.45}\endq
\begin{align}
&A_k(\bu,\ \lambda,\ \nu) = \cr
&{\Gamma(\bu+k)\over 2^{\bu+2}\pi^{2\bu}\Gamma(\bu)}\ 
(\bu+\lambda+\nu+2k)\times \cr
&{\Gamma(1+\lambda+\nu+2k)\Gamma(\bu+\lambda+k)\Gamma(\bu+\nu+k)
\Gamma(\bu+\lambda+\nu+k)\over
\Gamma(1+k)\Gamma(1+\lambda+k)\Gamma(1+\nu+k)\Gamma(1+\lambda+\nu+k)
\Gamma(1+\bu+\lambda+\nu+2k)}\ ,
\label{zm.50}\end{align}
\begin{align}
&B_k(\bu,\ \lambda,\ \nu,\ \z) = \cr
  &(1-\z^{-2})^{-{\bu\over 2}}
\Big [1+(1-\z^{-2})^\half \Big ]^{-\bu-\lambda-\nu}
\Big [\z + (\z^2-1)^\half \Big ]^{-2k} \times \cr
& F \left (  
\bu,\ 1-\bu\ ;\ \bu+1+\sigma\ ;\ 
{-\z + (\z^2-1)^\half \over 2(\z^2-1)^\half} \right )\ ,\ \
\sigma = \lambda+\nu+2k\ .
\label{zm.60}\end{align}
The conditions $(\bu,\ \lambda,\ \nu)\in V$, $\z\in \wh\Delta_1$, 
ensure that $S_k$ is well-defined for every $k\in \bZ_+$.
The main point of the proof is the convergence of $\sum_k S_k$. Once this is
established the analytic functions on both sides of
(\ref{zm.40}) are known to coincide on $U_0\times \Delta_2$, where
$U_0$ is defined by the conditions (\ref{x.14}), hence they coincide
wherever they are both defined.

For any $\k$, $\z \mapsto (1- \z^{-2})^\k$ is even
and analytic in $\wh \Delta_1$.
For $\z \in \Delta_1$,
$(\z^2-1)^\half = \z(1- \z^{-2})^\half$.
Let
\beq
z = {-\z + (\z^2-1)^\half \over 2(\z^2-1)^\half} =
{-1+(1-\z^{-2})^\half\over 2(1-\z^{-2})^\half}\ .
\label{zm.90}\endq
The map $\z \mapsto z$ is even and holomorphic in 
$\wh \Delta_1$, and it maps $\wh \Delta_1$ onto
$\{u\in \bC\ :\ \Re u < \half \}$.

For $\z \in  \Delta_1$, we set
\beq
\z = \ch(\omega),\ \  \omega = \tau+i\theta,\ \  \tau >0,\ \
|\theta| \le \pi.
\label{zm.70}\endq
Then
\beq
\Big [\z + (\z^2-1)^\half \Big ] = e^{\omega}\ ,
\label{zm.80}\endq
\beq
z = {1\over 1-e^{2\omega}},\ \ |z| \le {1\over e^{2\tau}-1}\ ,
\ \ \ 1+|z| \le {1\over 1-e^{-2\tau}}\ ,
\label{zm.100}\endq
\beq
(1-\z^{-2})^\half = {\sh(\omega)\over \ch(\omega)},\ \
\sh(\tau) \le |\sh(\omega)| \le \ch(\tau),\ \
\sh(\tau) \le |\ch(\omega)| \le \ch(\tau)\ .
\label{zm.101}\endq
In order to estimate the hypergeometric function appearing in (\ref{zm.60}),
we use \cite[2.3.2 pp 76-77]{bateman}\footnote{We use formula (11) p. 76
(as corrected in the Errata) in the case $n=0$.}
This provides the following estimate. Let
$a= \alpha+i\alpha'$, $b= \beta+i\beta'$, $c = \gamma+i\gamma'$,
$w\in \bC$, and suppose:
\beq
\beta \ge \half,\ \ \gamma-\beta \ge \half,\ \ 
|1-w| > \veps,\ \ |\arg(1-w)|< \pi(1-\veps)\ \ {\rm for\ some}\ \veps >0\ .
\label{zm.110}\endq
Then
\beq
|F(a,\ b,\ ;\ c\ ;\ w)| \le 1+ {|a||w|M(w)|\beta|
\sqrt{\ch (\pi\beta')\ch\pi(\gamma'-\beta')}\over
\gamma}\ ,
\label{zm.120}\endq
with
\beq
M(w) = \sup_{0\le u\le 1}|(1-uw)^{-a-1}|.
\label{zm.130}\endq
To apply this to the hypergeometric function appearing in (\ref{zm.60}),
we treat separately the cases $\Re \bu \ge \half$ and $\Re \bu \le \half$.

\noindent{\bf (1) Case $\Re \bu \ge \half$.}
In this case we set $a = 1-\bu$, $b = \bu$,
$c = \bu+1+\sigma = \bu+1+\lambda+\nu+2k$, and $w = z$ defined as above,
and satisfying (\ref{zm.100}) as well as $\Re z < \half$. For sufficiently
large $k$, the conditions (\ref {zm.110}) are satisfied and $\gamma > 1$.
We have $-\alpha-1-i\alpha' = \Re \bu -2+i\Im \bu$ and
\beq
M(z) \le \left \{ \begin{array}{l}
e^{2\pi|\Im \bu|}2^{2-\Re \bu}\ \ {\rm if\ } \Re \bu -2 < 0\ ,\\
e^{2\pi|\Im \bu|} (1+|z|)^{\Re \bu -2} \ \ {\rm if\ } \Re \bu -2  \ge 0\ .\\
\end{array}\right .
\label{zm.140}\endq
Hence, with (\ref{zm.70}) and (\ref{zm.100}),
\beq
|z| M(z) \le 4e^{2\pi|\Im \bu|}(e^{2\tau}-1)^{-1}(1-e^{-2\tau})^{-(\Re \bu-2)_+}\ ,
\label{zm.145}\endq
\begin{align}
& |B_k(\bu,\ \lambda,\ \nu,\ \z)| \le
e^{2\pi |\Im \bu|} 
\left ({\ch(\tau)\over\sh(\tau)} \right )^{\Re\bu} \times \cr
& e^{2\pi |\Im(\bu+\lambda+\nu)|}
\left ( {e^\tau\over \sh(\tau)}\right )^{-\Re(\bu+\lambda+\nu)_-}
\left ( {e^\tau\over \ch(\tau)}\right )^{-\Re(\bu+\lambda+\nu)_+}  \times \cr
&\Big [ 1+4|1-\bu|e^{2\pi|\Im \bu|}(e^{2\tau}-1)^{-1}(1-e^{-2\tau})^{-(\Re \bu-2)_+}
\ch(\pi\Im\bu)^\half\ch(\pi\Im \sigma)^\half \Big ]\times \cr
&\times e^{-2k\tau}\ .
\label{zm.150}\end{align}

\noindent {\bf (2) Case $\bu \le \half$.}
In this case we set $a = \bu$, $b = 1-\bu$,
$c = \bu+1+\sigma = \bu+1+\lambda+\nu+2k$, and $w = z$ defined as above.
For sufficiently
large $k$, the conditions (\ref {zm.110}) are satisfied and $\gamma > 1$.
We have $-\alpha-1-i\alpha' = -\Re \bu -1-i\Im \bu$ and
\beq
|z| M(z) \le 4e^{2\pi|\Im \bu|}(e^{2\tau}-1)^{-1}(1-e^{-2\tau})^{(\Re \bu+1)_-}\ ,
\label{zm.160}\endq
\begin{align}
& |B_k(\bu,\ \lambda,\ \nu,\ \z)| \le
e^{2\pi |\Im \bu|} \left ({\ch(\tau)\over\sh(\tau)} \right )^{|\Re\bu|}
\times \cr
& e^{2\pi |\Im(\bu+\lambda+\nu)|}
\left ( {e^\tau\over \sh(\tau)}\right )^{-\Re(\bu+\lambda+\nu)_-}
\left ( {e^\tau\over \ch(\tau)}\right )^{-\Re(\bu+\lambda+\nu)_+}  \times \cr
&\Big [ 1+4|\bu|e^{2\pi|\Im \bu|}(e^{2\tau}-1)^{-1}(1-e^{-2\tau})^{(\Re \bu+1)_-}
\ch(\pi\Im\bu)^\half\ch(\pi\Im (\sigma+2\bu))^\half \Big ]\times \cr
&\times e^{-2k\tau}\ .
\label{zm.170}\end{align}

The estimates (\ref{zm.150}) and (\ref{zm.170})
hold when $(\bu,\ \lambda,\ \nu)$ remains in a compact subset $K$
of $V$ intersected with $\{\Re \bu \ge \half\}$ or 
$\{\Re \bu \le \half\}$ respectively, and for all $k \ge k_0(K)$.
With the same restrictions on $(\bu,\ \lambda,\ \nu)$, 
we can estimate $A_k(\bu,\ \lambda,\ \nu)$ by the Erd\'elyi-Tricomi theorem
(\cite[5.11.13]{NIST}, \cite{T-E}, \cite[pp. 118 ff]{Olver}) and
we find that there is a $k_1(K) > k_0(K)$ such that, for every
$k \ge k_1(K)$,
\beq
|S_k| \le H \ch(\tau)^P \tau^{-Q} k^N e^{-2\tau k}\ ,
\label{zm.180}\endq
where $H >0$, $P>0$, $Q> 0$ and $N>0$  may depend on $K$ (but not on $\tau$).
Therefore $\sum_k S_k$ (resp. $\sum_{k=k_1}^\infty S_k$)
converges absolutely
and uniformly, when all the parameters remain in
$K\times \{\z = \ch(\tau+i\theta)\ :\ \veps < \tau < \veps^{-1}\}$,
 ($\veps > 0$), to a function
$h(\bu,\ \lambda,\ \nu,\ \z)$ (resp. $h_{k_1}(\bu,\ \lambda,\ \nu,\ \z)$)
holomorphic in these variables. Since $\sum_{k=0}^L S_k$ is holomorphic in
$\wh \Delta_1$ it follows, by the maximum principle, that the convergence
extends to $K\times \{\z = \ch(\tau+i\theta)\ :\ \veps < \tau \}\cup \{\infty\}$,
i.e. the functions $h$ and $h_{k_1}$ are holomorphic
there.

For any function $f$ holomorphic in $\Delta_1$, and $P>0$, $Q>0$, let
\beq
||f||_{P,Q} = \sup_{\z\in \Delta_2} e^{-P\tau} \tau^{Q} |f(\z)|,\ \ \ 
\z= \ch(\tau+i\theta),\ \ \tau >0,\ \ |\theta| < \pi\ .
\label{zm.190}\endq
It then follows from our estimates that if $(\bu,\ \lambda,\ \nu)$
remains in the compact $K$, there exist constants $P>0$, $Q>0$, and $M>0$
such that
\beq
||\sum_{k=k_1}^L S_k ||_{P,Q} \le M \ \ \ \forall L > 0,
\label{zm.200}\endq
for any $\veps >0$ there is an $L>0$ such that
\beq
||\sum_{k=L}^\infty S_k ||_{P,Q} < \veps\ ,
\label{zm.210}\endq
\beq
||h_{k_1} ||_{P,Q} \le M \ .
\label{zm.220}\endq
(The functions appearing in the lhs of these inequalities are considered
as functions of $\z\in \Delta_2$ at fixed $\bu$, $\lambda$, $\nu$.)
Restricting these estimates to the tuboid $\ZZ_{1-}\times \ZZ_{1+}$
it is possible to show that (\ref{zm.40}) remains true for the
boundary values of both sides in the sense of tempered distributions.
We omit the lengthy but straightforward
details.

\end{document}